\documentclass[aps,twocolumn,superscriptaddress,groupedaddress,nofootinbib]{revtex4}
\usepackage[utf8]{inputenc}
\usepackage{amsmath}
\usepackage{amssymb}
\usepackage{dsfont}
\usepackage{graphicx}
\graphicspath{{figs/}}
\usepackage{soul}

\usepackage{hyperref}
\hypersetup{
    colorlinks,
    linkcolor={blue},
    citecolor={blue},
    urlcolor={blue}
}

\usepackage{comment}

\usepackage{pythonhighlight}
\definecolor{deepblue}{rgb}{0,0,0.5}
\lstnewenvironment{pythonLines}[1][]
{\lstset{style=mypython,
emph={DensePauliExpectation},
emphstyle=\color{deepblue}}}{}

\newcommand{\Com}{\text{Com}}

\begin{document}

\title{Fast Partitioning of Pauli Strings into Commuting Families\\
for Optimal Expectation Value Measurements of Dense Operators}

\author{Ben Reggio\footnote{breggio2@illinois.edu}}
\author{Nouman Butt\footnote{ntbutt@illinois.edu}}
\author{Andrew Lytle\footnote{atlytle@illinois.edu}}
\author{Patrick Draper\footnote{pdraper@illinois.edu}}

\affiliation{
Illinois Quantum Information Science and Technology Center, Urbana, IL 61801\\
Illinois Center for Advanced Studies of the Universe, Urbana, IL 61801\\
Department of Physics, University of Illinois at Urbana-Champaign, Urbana, IL 61801}

\begin{abstract}
    
    The Pauli strings appearing in the decomposition of an operator can be can be grouped into commuting families,  reducing the  number of quantum circuits needed to measure the expectation value of the operator. We detail an algorithm to completely partition the full set of Pauli strings acting on any number of qubits into the minimal number of sets of commuting families, and we provide python code to perform  the partitioning. The partitioning method scales linearly with the size of the set of Pauli strings and it naturally provides a fast method of diagonalizing the commuting families with quantum gates. We provide a package that integrates the partitioning into Qiskit, and use this to benchmark the algorithm with dense Hamiltonians, such as those that arise in matrix quantum mechanics models, on IBM hardware. We demonstrate computational speedups close to the theoretical limit of $(3/2)^m$ relative to qubit-wise commuting groupings, for $m=2,\dotsc,6$ qubits.
\end{abstract}

\maketitle

\section{Introduction}
\label{sec:Introduction}
The cost of a quantum computation depends on several aspects of the computation, including the number of required quantum circuits, the depth of the circuits, and the number of times the same circuits have to be run in order to achieve a level of confidence in the results. There are also classical preprocessing costs that depend on the algorithms used to generate circuits for execution on a quantum device. For computations involving expectation value measurements, e.g.\ variational quantum eigensolver (VQE) problems, the na\"ive approach for a generic operator produces of order $4^m$ circuits for $m$ qubits.   In the NISQ era, the capacity to share the computational burden between classical and quantum computers in an optimal way will be crucial.  

In this paper we revisit the problem of partitioning $m$-qubit Pauli strings into commuting families, with the goal of developing a practical implementation of a complete solution. As we will describe further below, our implementation is optimal for problems where an order one fraction of the full set of Pauli strings need to be grouped. Pauli partitioning is a classical problem, the solution of which can be used to reduce the number of circuits needed to measure an expectation value on a quantum device. For example, to optimally characterize the density matrix describing an $N=2^m$-state system by repeated measurements from an ensemble of identically prepared states, one may expand the density matrix in a basis of  $N^2 -1$  Pauli strings. In the brute force approach one must then measure all of the strings. Partitioning of the Paulis into $N+1$ commuting sets of size $N-1$ can reduce the required number of  measurements by a square root. Although the scaling with the number of qubits is still exponential, in the NISQ era such improvements can be of great value:  a complete characterization of, say, a 7-qubit state is currently feasible only with grouping. For another example, expectation value measurements of observables with dense subspaces (e.g. a nearly-block-diagonal Hamiltonian, with small dense diagonal blocks and sparse off-diagonal blocks) complete partitioning of the subspace operators is again desirable.

The problem of partitioning Pauli strings has an interesting history, and many authors have contributed to developments and applications. Let us give a brief and necessarily incomplete survey. In terms of practical implementations the idea has been studied by a number of authors~ (see, for example, \cite{yen2020measuring,Gokhale2019MinimizingSP,Huggins_2021} and references therein) and has been shown to lead to considerable advantage for quantum chemistry problems. From a theoretical perspective, the problem and its  relatives have been solved with different methods in the mathematics and QIS literature. An early characterization of the problem is known in the literature as the construction of mutually unbiased bases (MUBs)~\cite{WOOTTERS1989363}. Wootters and Fields provide an explicit construction of $N+1$ MUBs for unique determination of the state through $N+1$ measurements. An optimal solution to the Pauli partitioning problem also provides a solution for a complete set of MUBs. As described above, this solution can be used to reduce the exponential scaling of na\"ive measurements by a square root, and can be instrumental for state-tomography methods. In the mathematical literature, the problem of Pauli partitioning is related to the orthogonal decomposition of Lie alebgras of type $A_{n -1}$~\cite{kostrikin2011orthogonal} into a direct sum of Cartan subalgebras (CSA), where each CSA is pairwise orthogonal with respect to the Killing form. An explicit construction for $\mathfrak{sl}(n,\mathcal{C})$ with $n=4$ can be found in~\cite{sriwongsa2019orthogonal}. The orthogonal decomposition approach is equivalent to finding a set of MUBs~\cite{boykin2005mutually}. We also note that the construction of MUBs has potential applications to quantum key distribution protocols, where they provide increased tolerance to noise and maximize the value of the secret key rate~\cite{boykin2004information,Bru__1998,PhysRevLett.88.127902,PhysRevLett.85.3313}.

Most relevant for our study are presented here~\cite{Gokhale2019MinimizingSP, Kern_2010, jena_paper, jena_2019,sarkar2019sets}. Ref.~\cite{Gokhale2019MinimizingSP} uses graph-theoretic methods to construct a partitioning. This approach is particularly useful in problems involving expectation values of sparse observables, but for the dense case a complete solution is essential. In Qiskit~\cite{Qiskit}, the AbelianGrouper routine uses a graph-theoretic approach to partition strings into $O(N^\frac{3}{2})$ families. Our approach shares the most in common with Ref. \cite{Kern_2010}, which has constructed solutions up to $m=24$ qubits, and with Refs. ~\cite{jena_2019,jena_paper}, which proved the existence of an algorithm to fully partition Pauli strings for any $m$ using Singer cycles and companion matrices.

The novel contributions of our work are primarily of a practical nature. We begin by describing a complete algorithm to sort the full set of Pauli strings, for any $m$, into the minimum number of families. This algorithm is essentially the same as what appears in \cite{Kern_2010,jena_2019,jena_paper}, but we describe it in elementary linear algebra terms using a minimum of mathematical machinery, and we give an efficient function that maps any input string to its family. We further extend the algorithm by giving an explicit construction of the unitary operators needed to rotate each family into the computational basis. We also describe and provide a new publicly-available Qiskit~\cite{Qiskit} module that implements the algorithms, generating a minimal partition and the corresponding unitaries for measurement of each partition. Finally, we benchmark the module on simulations of models arising in high energy physics, and we also benchmark the performance against the grouping strategies currently implemented in Qiskit.
We demonstrate speedups in runtime on quantum hardware close to that implied by counting groups, and we demonstrate favorable scaling of the classical resources needed to generate solutions, compared with graph theory-based methods.

The rest of this work is organized as follows. In Section \ref{sec: Properties} we review basic properties of Pauli strings. Section \ref{sec: Constructing Perfect solutions} describes the considerations and tools needed to construct a perfect solution, or a complete partitioning  of all strings into a minimal number of families, and  Section \ref{sec: Algorithmic construction} describes the solution-generating algorithm in detail. Readers interested only in the description of the code and benchmarking studies may wish to skip to Section \ref{sec: Implementation}, where we present benchmark results for the algorithm on IBM quantum devices and compare the computational cost 
 of circuit runtimes with existing methods. We also show measurement accuracy in relation to other grouping strategies. In Section~\ref{sec:vqe}  we apply it to a simulation of a physical model related to the QCD vacuum with a non-trivial hamiltonian and demonstrate overall speedup with increasing number of qubits using timings of VQE runs. In Section~\ref{sec:graphcomparison} we provide a comparison of computational efficiency and accuracy against other grouping strategies based on graph-theoretic methods.  Section \ref{sec:code} describes the public code packages developed using the method described in section \ref{sec: Constructing Perfect solutions} and \ref{sec: Algorithmic construction}. The code generates quantum circuits that can be run on any hardware that implements Clifford gates. In section \ref{sec: conclusion} we summarize our results. 

Additionally, Appendices describe the algorithm from Sec. \ref{sec: Algorithmic construction} in more detail. Appendix \ref{sec:matrix_algorithms} shows the matrix algebra which proves the validity of the groupings, and Appendix \ref{sec:circuit_rep} describes how the change of basis circuits are represented.

\section{Properties of Pauli Strings}\label{sec: Properties}

In this section we discuss some of the basic properties of Pauli strings and commuting families.  These properties provide scaffolding for the subsequent development of a constructive algorithm to sort strings into commuting families. 

A Pauli string is a tensor product of $m$ factors of four types of 2 $\times$ 2 matrices: the identity matrix and the three Pauli matrices. It may be represented as a matrix, a tensor product, or a string. For example, the tensor product $\sigma_x \otimes I \otimes \sigma_z$ in the string representation is $XIZ$, and a matrix representation of the same string in the standard basis is:
\begin{equation}
    XIZ = \sigma_x \otimes I \otimes \sigma_z =\begin{pmatrix}
        0 & 0 & 0 & 0 & 1 & 0 & 0 & 0 \\
        0 & 0 & 0 & 0 & 0 & -1 & 0 & 0 \\
        0 & 0 & 0 & 0 & 0 & 0 & 1 & 0 \\
        0 & 0 & 0 & 0 & 0 & 0 & 0 & -1 \\
        1 & 0 & 0 & 0 & 0 & 0 & 0 & 0 \\
        0 & -1 & 0 & 0 & 0 & 0 & 0 & 0 \\
        0 & 0 & 1 & 0 & 0 & 0 & 0 & 0 \\
        0 & 0 & 0 & -1 & 0 & 0 & 0 & 0
    \end{pmatrix}
\end{equation}

We denote the length of the string representation, by $m$, which is the number of factors in the tensor product and also the number of qubits. The size of the matrix is $N \times N$, where $N = 2^m$. $N$ and $m$ will continue to refer to these properties of the Pauli strings. Two Pauli strings of the same length must either commute or anticommute, since they are tensor products of Pauli matrices which have the same property. They square to the identity for the same reason.

A ``family'' is defined as a maximally populated  set of commuting strings. More formally we define a family $f$ as a set of strings such that every string commutes with every other string, and there exist no outside strings which commute with all members of the family. The identity matrix could exist in every family, so we ignore it until the rest of the strings are partitioned. We have noted that a family is a Cartan subalgebra of the $\mathfrak{su}(N)$ algebra. It is known that these consist of $N-1$ elements. We will show this at the end of the section.

We define a ``generating set of strings'' as a set of strings (or ``generators,'' now that they are associated with a generating set) that has the properties that no generator is a product of the other generators, and all of the generators commute. If $[A,C]=[B,C]=0$, then 
\begin{align}
    [AB,C] = [A,C]B + A[B,C] = 0
\end{align}

$AB$ commutes with anything that commutes with $A$ and $B$, so all the elements of a family containing $A$ and $B$ commute with $AB$, and $AB$ must be included in the family by definition. This excludes the identity, but is a group associated with each family that includes the identity, which is closed under matrix multiplication. Every product of the generators is unique and commuting. We will show an inductive method to get a set of generators from any family. 

Suppose there is a set of $k$ generating strings $\{S_1, \dotsc, S_k \}$. Construct a set of commuting strings $g_k$ from the products of the generators, including the generators themselves. Assume that all $B \in g_k$ can be written uniquely as $B = \prod_{i=1}^k S_i^{b_i}$ for a vector $b \in \mathbb{Z}_2^k$. There are $2^k - 1$ vectors in $\mathbb{Z}_2^k$, so the cardinality of $g_k$ is $2^k-1$.

We are going to define a process that we call an "extension" where we want to add a string into this set of generating strings. When we add a commuting generator which is not in $g_k$, there is a larger set $g_{k+1}$ which includes products of this new set of generators. This new set should have cardinality $2^{k+1} - 1$, and also have the property that each product of generators is unique. When we define this process, we can use induction to show that all sets of generating strings produce sets of commuting strings of length $2^k - 1$, where $k$ is the number of generators. 

If there is a string $S_{k+1}$ such that $S_{k+1} \notin g_k$ and $S_{k+1}$ commutes with all the generators of $g_k$, $\forall i \leq k: [S_{k+1}, S_i] = 0$. 
Then we can prove by contradiction that
\begin{equation}
    \forall B \in g_k: S_{k+1} B \notin g_k
\end{equation}
To do this, assume that 

\begin{align}
    \exists B \in g_k: S_{k+1} B \in g_k \\
    \exists b \in \mathbb{Z}_2^k: B = \prod_{i=1}^k S_i^{b_i} \\
    \exists a \in \mathbb{Z}_2^k: S_{k+1} B = \prod_{i=1}^k S_i^{a_i} \\
    (S_{k+1} B) B = S_{k+1} = \prod_{i=1}^k S_i^{a_i + b_i} = \prod_{i=1}^k S_i^{c_i} \,,
\end{align}
where addition of $a$ and $b$ is defined mod 2. $S_{k+1}$ can be written as $\prod_{i=1}^k S_i^{c_i}$ where $c_i \in \mathbb{Z}_2^k$, so $S_{k+1} \in g_k$. This is a contradiction, so any product $S_{k+1} B \notin g_k$ for all $B$ in $g_k$. Since Pauli strings are invertible, each product is unique for each unique $B$.

The new set $g_{k+1}=\{ g_k, S_{k+1}, S_{k+1} g_k \}$ has cardinality $2^{k+1} -1$. $g_{k+1}$ is generated by $\{S_1, \dotsc, S_{k+1} \}$, i.e. all elements can be written uniquely as $\prod_{i=1}^{k+1} S_i^{b_i}$ for $b_i \in \mathbb{Z}_2^{k+1}$. Since all the generators of the new set commute, all of the elements of the new set $g_{k+1}$ commute. Following this induction, any commuting set $g_k$ with $k$ generators can be extended to another commuting set $g_{k+1}$ with $k+1$ generators if there is a string outside $g_k$ which commutes with all of the generators of $g_k$. The induction may begin with the set of a single string.

We may also use these properties to select, from any family $f$, a set of $m$ generators. First, select any string $S_1 \in f$. Define the set $g_1 = \{ S_1 \}$. $g_1$ is generated by $S_1$. We will inductively extend $g_1 \rightarrow g_2$, $g_2 \rightarrow g_3$, etc.\ using the previous process until we find $g_m=f$. Suppose we have reached a point where $g_k \subseteq f$ has cardinality $2^k -1$, and generators $\{S_1, \cdots, S_k\}$. Select any string $S_{k+1}$ such that $S_{k+1} \in f \setminus g_k$. Extend $g_k$ to the set $g_{k+1} = \{ g_k, S_{k+1}, S_{k+1} g_k \}$. Any string in $S_{k+1} g_k$ is in $f$, since all such strings commute with every string in $f$, and $f$ is assumed to be maximal. The process of adding generators can be repeated until $f = g_m$, and the generators of $g_m$ provide a set of generators of $f$. The simplest example of a family and a set of generators is the {\emph{$z$ family}}, which we will also refer to as the {\emph{diagonal family}}, with generators $\{I...IIZ, I...IZI, ...\}$. Any string which contains an $X$ or $Y$ automatically does not commute with the corresponding generator with a $Z$ in that position, so this family contains only strings with $Z$ and $I$, of which there are $2^m - 1$.

There is also a way to define a map between the families using a string. Consider the transformation $U_{i}\equiv \exp \big(i \frac{\pi}{4} P_i \big)$, where $P_i$ is a Pauli string. Any Pauli string $P_j$ transforms to another Pauli string under $P_j\rightarrow U_i P_j U_i^\dagger$. Since it is a unitary transformation, commutativity is preserved, so transforming a family produces another family. The transformation has the following property:
\begin{equation}
    U_i P_j U_i^\dagger = \left\{ \begin{array}{cc}
        P_j & \text{if } [P_i , P_j] = 0 \\
        
        i P_i P_j & \text{if } [P_i , P_j] \neq 0
    \end{array} \right\}
\end{equation}

In Section \ref{sec: Algorithmic construction}, we use this transformation to show that any family can be transformed into any other family by finding an appropriate set of transforming strings $\{ P_{i_1} ... P_{i_n} \}$, so every family has the same cardinality as the diagonal family. There are $m$ generators for this family defined above, and any tensor product of $\sigma_z$ and $I$ can be written as a product of these generators. Every family has cardinality $2^m - 1$.

We define a {\emph{perfect solution}} to be a partitioning of all $4^m-1$ Pauli strings on $m$ qubits into $2^m+1$ families. In the next section we detail a constructive algorithm to produce perfect solutions.

\section{Constructing Perfect Solutions} \label{sec: Constructing Perfect solutions}

\subsection{Preliminaries}
\label{subsec:Preliminaries}

In this section, we begin the construction of perfect solutions from product tables of strings and find certain conditions on the construction. These conditions will lead us to express families in terms of $\mathbb{Z}_2$ valued matrices.
 
We begin by selecting two arbitrary families. We consider a canonical pair to be the diagonal $z$ family, $\{z_1, z_2, \dotsc, z_{N-1} \}$ which is the family which contains strings of the characters $I$ and $Z$, along with the $x$ family $\{x_1, x_2, \dotsc, x_{N-1} \}$ which is constructed similarly but with the $X$ character instead of $Z$.
\begin{table}[ht]
    \centering
    \begin{tabular}{c|c c c}
         & $IX$ & $XI$ & $XX$ \\
         \hline
         $IZ$ & \fbox{$IY$} & \underline{$XZ$} & \textbf{XY}\\
         $ZI$ & \textbf{ZX} & \fbox{$YI$} & \underline{$YX$} \\
         $ZZ$ & \underline{$ZY$} & \textbf{YZ} & \fbox{$YY$}
    \end{tabular}
    \caption{Example of a solution table for $m=2$. The $x$ and $z$ families are used to build the inner columns and rows. One family is boxed, one is underlined, and the third shown in bold. Each family contains exactly one string from each inner row and column.}
    \label{tab:product table}
\end{table}

Construct a table with the members of $z$ in the far left column, and the members of $x$ in the top row. The inner members of the table are the matrix products of the member of $z$ in their row and the member of $x$ in their column. For now, we may ignore phases that appear in these products. Define the string in the $i$th row and the $j$th column to be $S_{i,j}$. An example of such a product table is shown for 2 qubits in Table \ref{tab:product table}.

There are $2N-2$ strings in the $x$ and $z$ families combined, and $(N-1)^2 = N^2 - 2N + 1$ strings represented by $S_{i,j}$, for a total of $N^2 - 1$  labels. They are unique because the product table of $\{I,Z\}$ and $\{I,X\}$ contains each Pauli matrix and the identity exactly once. All of the $N^2 - 1$ strings appear exactly once in the product table of the $z$ and $x$ families.

Each family in a perfect solution, apart from the $x$ and $z$ families, contains exactly one string from each interior row and column of the product table. If a family contained two strings from the same row or column, then it would also contain the product of the two. If the strings $S_{i,j}$ and $S_{i,k}$ are in a family, the family would also contain
\begin{align}
    z_i x_j z_i x_k = \pm x_j z_i z_i x_k = \pm x_j x_k \in x
\end{align}
The result is already in the $x$ family (or the $z$ family, if two strings in the same column are in the same family). This implies that the string is repeated and the solution is not perfect. We conclude that no more than one string can be picked from each row or column. There are $N-1$ rows and columns, and each family has $N-1$ strings, so each family in a perfect solution must contain exactly one string from each row and column. 

\subsection{Latin Square Formulation}
At this stage, it is natural to think of selecting a solution as analogous to filling out a Latin Square. A Latin square is an $(N-1) \times (N-1)$ grid of numbers where each number appears exactly once in each row or column. It is a good exercise to try solving the problem this way. Use the rows of the Latin square to indicate the families, the columns to represent which string from the $z$ family is being multiplied, and the number in the grid to indicate the string from the $x$ family being multiplied. An example of this is shown in Table \ref{tab:latinsquare}. It is clear that any perfect solution will have an associated Latin Square, and following this line of inquiry further will give us some insight into how to construct perfect solutions, but we will see it does not quite automate the task.
\begin{table}
    \centering
    \begin{tabular}{|c|c|c|}
        \hline
        1 & 2 & 3 \\
        \hline
        2 & 3 & 1 \\
        \hline
        3 & 1 & 2 \\
        \hline
    \end{tabular} $\rightarrow$
    \begin{tabular}{|c c c|}
        \hline
        $IY$ & $YI$ & $YY$ \\
        \hline
        $XZ$ & $YX$ & $ZY$ \\
        \hline
        $XY$ & $ZX$ & $YZ$ \\
        \hline
    \end{tabular}
    \caption{Latin Square formulation of a perfect solution. The Latin Square on the left corresponds to the solution on the right. For each box on the left, take the $z$ string corresponding to the column, and multiply it by the $x$ string corresponding to the entry in that box. For the middle box, the row is 2 and the entry is 3, so multiply $z_2 = ZI$ (think binary) times $x_3 = XX$ to get $YX$, which is seen in the same position on the right, in the family $XZ, YX,  ZY$.}
    \label{tab:latinsquare}
\end{table}

An obvious deficiency is there is no restriction in the Latin Square which prevents non-commuting strings from being placed in a family, so every time one places a number in the Latin Square, one must first check whether the string corresponding to this number commutes with the rest of the strings in the family. Since the families are associated with a group closed under multiplication, any products between old strings and new strings must be filled in. 

Upon filling in several squares with solutions, patterns begin to emerge. The first pattern emerges when we begin with a ``canonical" family, which is the row of ordered integers 1,2,3... The first column may also be organized canonically, so that the first column also reads the ordered integers 1,2,3...  We observe that if the square is begun in this way, solutions correspond to symmetric Latin squares. Using this formulation, it is unclear why this is the case, which is a clue that there is more to discover.

Every time a single number is placed in the Latin square, an entire block is filled out using closure under multiplication and symmetry. It becomes clear that only the \emph{generating} rows and columns need to be filled in, and the rest will follow because of the extension defined in the previous section. When filling out Table \ref{tab:latinsquare}, the first row and column are trivial. To pick the middle square, the only strategy is to guess and check possible values. If the guess is 3, this corresponds to the generating string $YX$. Our previous section shows that if $YX$ is in a family with $XZ$ (which is already in this row/family), then their product $ZY$ must also be in this family, which corresponds to the 1 at the end of the row of the square. Every new string we add into the Latin square through guess and check can be treated as a generating string, and exponentially more squares can be filled in using the group extension for every generating string. This is a another hint that there may be a another approach that is more compact, and only uses the generating strings.

We have remarked above that commutation has to be checked by hand. Another task for the solution builder is to avoid contradictions in the Latin square. Contradictions arise when filling in a block forces one to repeat a number in a row or column. In some cases there are no valid options for a position in the grid. It is unclear at this stage if one can  predict when such contradictions will arise without explicitly working out the consequences of adding a string to a family.

To resolve these issues, we will take a somewhat different approach which makes it more straightforward to impose commutativity and the uniqueness of entries in rows and columns.

\subsection{Commutativity and a Binary Encoding}
We first consider the issue of string commutativity and rewrite it in a useful way as a problem in binary arithmetic. We see in this section that encoding the strings using vectors over $\mathbb{Z}_2$  allows for a convenient expression of the commutativity condition between strings. Eventually, we will express the families in a solution in terms of $\mathbb{Z}_2$-valued matrices, and the commutativity formula obtained in this section is useful in the construction.

The commutator of two strings $S_{i,j}= z_ix_j$ and $S_{k,l}=z_k x_l$ is
\begin{align}
\label{eq: CEq1}
    [z_i x_j, z_k x_l] = \left\{ \begin{array}{cc}
        z_i z_k x_j x_l - z_k x_l z_i x_j,  &~~ [x_j, z_k] = 0 \\
        -z_i z_k x_j x_l - z_k x_l z_i x_j,  &~~ \{x_j, z_k\} = 0 \\
    \end{array} \right\}
\end{align}

Every Pauli string either commutes or anticommutes with every other string.
The commutation of  two strings $S_{i,j} = z_{i} x_{j}$ and $S_{k,l} = z_k x_l $ only depends on  whether $z_i$ commutes with $x_l$, and whether $z_k$ commutes with $x_j$. Define a map $\Com(x_i,z_j)$ from an $x$ string and a $z$ string to $\{0,1\}$, which maps to 0 if the strings don't commute, and 1 if the strings do commute. It is straightforward to check that $S_{i,j}$ and $S_{k,l}$ commute if $\Com(x_j,z_k) = \Com(x_l,z_i) $:
\begin{align}
 \label{eq: CEq2}
    [z_i x_j, z_k x_l] = \left\{ \begin{array}{cc}
        0  & \Com(x_j,z_k) = \Com(x_l,z_i) \\
        - 2 z_k x_l z_i x_j  & \Com(x_j,z_k) \neq \Com(x_l,z_i) \\
    \end{array} \right\}.
\end{align}

The way to evaluate the $\Com$() map is to examine the paired characters at each position of the two strings, and for each time $X$ and $Z$ both appear in the same position, the value of the commutator switches between 0 and the product of the two strings. This corresponds to $\mathbb{Z}_2$ arithmetic.

It is now convenient to adopt a binary encoding of the strings. Strings in the $z$ family have only two possible characters for each of $m$ positions, which can be mapped to a binary representation. Let 0 correspond to an $I$ and 1 to an $X$ or $Z$ depending on the family. Define the strings $x_i$ and $z_i$, $i=1\dots N-1$, by the binary representation of the label $i$. For example, for $m=4$, $x_{15} = XXXX$ and $z_{3} = IIZZ$. We also define $m$-dimensional vectors $v_i$ over $\mathbb{Z}_2$ corresponding to the binary representation of the integer $i$: $v_i$ is created by converting $i$ to binary, and populating the vector element $(v_i)_j$ with the $j$th digit of $i$.  A consequence of this definition is that $v_i + v_j = v_{i \oplus j}$ where $\oplus$ denotes $\mathbb{Z}_2$ addition. The vector space contains the vectors $v_i$ for $0 < i < N$ and the zero vector. For example, in the space corresponding to $m$ = 3,
\begin{equation}
    v_5 + v_6 = \begin{pmatrix} 
    1 \\
    0 \\
    1
    \end{pmatrix} +  \begin{pmatrix} 
    1 \\
    1 \\
    0
    \end{pmatrix} =  \begin{pmatrix} 
    0 \\
    1 \\
    1
    \end{pmatrix} = v_3 .
\end{equation}

As described above we can associate each vector $v_i$ with a string in the $x$ or $z$ family. A computationally valuable aspect of this encoding is that the commutation function $\Com()$ acting on a $z$ string and an $x$ string can be rewritten in terms of the corresponding $v$ vectors: 
\begin{equation}
    \Com(z_i, x_j) = f(i,j) \equiv \sum_k (v_i)_k (v_j)_k .
    \label{eq:usefuleq}
\end{equation}
The binary computation of the commutation function will be convenient for developing the algorithm to construct families.  Cf. Eq.~\ref{eq: CEq2}, the input to $f$ is the $v$-vector corresponding to the $z$-factor in the $xz$ decomposition of one string, and the $v$ vector corresponding to the $x$-factor of $xz$ decomposition of the other string. 

\subsection{Generating Matrix}
We have learned that a solution involves picking sets of generators for the families. We can pick a canonical set of generators for the $z$ family corresponding to powers of two, $z_{2^a}$ (i.e.\ the set of strings with a single $Z$,
$\{I \dotsb IIZ, I \dotsb IZI, \dotsc\}$). Along with this canonical set, for any family, there is a corresponding set of generators from the $x$ family which can be used to build the family's generators by multiplying the $x$ and $z$ generators. The rest of the strings in the family can be built from the products of the generators. Interestingly, this process has an analog in $\mathbb{Z}_2$ matrix multiplication. The linearity of the matrix multiplication corresponds to the closure under multiplication, since addition in $\mathbb{Z}_2$ corresponds to multiplication of strings.

Each family in a perfect solution containing the $z$ and $x$ families must contain exactly one string from each row and column in the product table of the $z$ and $x$ families, so perfect solutions of this type should be expressible as a certain permutation $P(i)$ acting on the integers $i=1\dots N-1$. For a family $F$, with a corresponding permutation $P$:
\begin{equation}
    \forall i, 0 \leq i < N:  z_i x_{P(i)} \in F
\end{equation}
Since families are associated with groups closed under multiplication, it is also true that the permutation is linear:
\begin{align}
    \exists i,j, i \neq j: \{ z_i x_{P(i)}, z_j x_{P(j)} \}  \subset F \\
    (z_i x_{P(i)}) \cdot (z_j x_{P(j)})) \in F \\
    z_{i \oplus j} x_{P(i) \oplus P(j)} \in F \\
    P(i \oplus j) = P(i) \oplus P(j)
\end{align}
This motivates a $\mathbb{Z}_2$-valued matrix definition of the permutation:
\begin{equation}
    A v_i = v_{P(i)}.
    \label{avvp}
\end{equation}
 
Therefore, associated with any family which may appear in a solution with the $x$ and $z$ families is an invertible $\mathbb{Z}_2$ valued matrix $A$. (Since a permutation is invertible, $A$ is invertible.) Since A is invertible, it defines a map from any complete basis in $\mathbb{Z}_2^m$ to another complete basis. This corresponds to a map from a generating set of $x$ strings to another generating set of strings. This is essentially what the Latin square does: it maps which generating $x$ strings go in which positions, and the rest of the strings are placed based on the location of the generators. Similar to the Latin Square, extensions can be used to find the rest of the strings.
 
In the previous section, we saw that the Latin square approach did not automatically enforce commutativity, and that one could encounter contradictions when filling it in. It turns out that commutativity and avoidance of contradictions can be enforced automatically by endowing $A$ with two more properties. 

The first condition is that $A$ must be symmetric. This arises as follows. Suppose we have a permutation $P$ acting on the first $N-1$ positive integers. We can identify a simple set of $m$ generating strings among the elements $x_i z_{P(i)}$ of the product table. Consider the strings $z_{2^a} x_{P(2^a)}$ and $z_{2^b} x_{P(2^b)}$, with $a \neq b$ and $0\leq a,b <m$. $2^a \oplus 2^b = 2^c$ is impossible, so these strings are generators. If they commute, then the rest of the family commutes. They commute if $f(2^a, P(2^b)) = f(2^b, P(2^a))$, and this condition implies symmetry of the $A$ matrix associated with the permutation:
\begin{align}
    f(2^a, P(2^b))&=\sum_c (v_{2^a})_c (v_{P(2^b)})_c\nonumber\\
    &=\sum_c (v_{2^a})_c(A v_{2^b})_c\nonumber\\
    &=\sum_{c,d} (v_{2^a})_c A_{cd} (v_{2^b})_d\nonumber\\
    &=\sum_{c,d} \delta_{ac}A_{cd} \delta_{b,d}\nonumber\\
    &=A_{ab}
    \label{symmetryeq}
\end{align}
We have used $(v_{2^a})_b = \delta_{ab}$. The condition on commutation is the symmetry of $A$:
\begin{align}
    f(2^a, P(2^b))= f(2^b, P(2^a))\Longleftrightarrow A_{ab}=A_{ba}.
\end{align}

The second constraint of a perfect solution is that no two families contain the same string. If one family defined by $A_i$ and another family defined by $A_j$ contain the same string, then:
\begin{align}
    \exists k: A_i v_k = A_j v_k \\
    (A_i - A_j) v_k = 0.
\end{align}
In this case $A_i - A_j$ must not be invertible. If $A_i - A_j$ is invertible, then the two families do not share a string. A perfect solution can then be expressed as a set of $N - 1,  \mathbb{Z}_2$ valued matrices $\{ A_1 ... A_{N-1} \}$ that are $m$ dimensional. They satisfy:
\begin{itemize}
    \item $A_i$ is symmetric
    \item $A_i - A_j$ is invertible \cite{jena_2019}
\end{itemize}
The solution so obtained also contains the canonical families, giving $N+1$ total families.

To summarize, the two conditions listed above encode the restrictions on the families that we set out to achieve. The symmetry property enforces commutativity within each family, and the invertibility property is associated with the uniqueness of the rows and columns of the Latin Square. We have specified that we are operating on qubits, so we use the field $\mathbb{Z}_2$. There are more general quantum systems using qudits, which have similar conditions for fields which are larger prime integers. In these cases, we refer to \cite{jena_2019} for the conditions necessary to create similar matrices in other finite fields.

\section{Algorithmic Construction of the generating Matrix and Solutions} \label{sec: Algorithmic construction}

\subsection{Symmetrizing Companion Matrices}
We need a method to generate a set of $A$-matrices with these symmetry and invertibility properties. There will be a similar method for other general $\mathbb{Z}_p$ for prime $p$, but the method for $\mathbb{Z}_2$ is slightly different. This section focuses specifically on qubit operators, so all matrices are $\mathbb{Z}_2$ valued.

One suitable choice to fulfill the invertibility condition is a $m \times m$ companion matrix ($C$), which was considered by Jena \cite{jena_2019}. (See Appendix~\ref{sec:matrix_algorithms} for the definition of a companion matrix and details of results used in this section.) The relevant property of $C$ is that it generates a permutation with one cycle and no fixed points. Such a cycle must have a period of $N-1$ points, and $C^i + C^j$ is always invertible for $i - j \neq 0 \text{ mod } (N-1)$.

The set $\{ C, C^2 ... C^{N-1} \}$ fulfills the invertibility condition, but not the symmetry condition. If there is an invertible matrix $B$ such that $B^{-1}CB = A$ is symmetric, then $A^i - A^j= B^{-1}(C^i - C^j) B$ is invertible:
\begin{equation}
    (A^i - A^j)^{-1} = B^{-1} (C^i - C^j)^{-1} B .
\end{equation}
There is an algorithm to construct the relevant similarity transformation. First, one can construct a matrix $D$ such that $DC^T = CD$  \cite{finitefields}. Now suppose that $D$ can be written as $D=B B^T$ for some invertible $B$. Then $A = B^{-1} C B$ must be symmetric:
\begin{align}
    C^T = B^{-T} B^{-1} C B B^T \\
    (B^{-1} C B)^{T} = B^{-1} C B \\
    A^T = A
\end{align}
There is an algorithm to explicitly construct $D$ and $B$. For any matrix $M$ with all diagonal elements equal to unity, one can find an invertible matrix $L$ so that $L L^T = M$, using this algorithm. It also assumes that the field is $\mathbb{Z}_2$. There are similar algorithms for other finite fields. We must first find a matrix $\Lambda$ such that $M \equiv \Lambda^T D \Lambda$ has diagonal elements equal to unity. This can be done algorithmically and produces an invertible $\Lambda$. Then:
\begin{align}
    \Lambda^T D \Lambda = M = LL^T
\end{align}
so
\begin{align}
    D &= \Lambda^{-T} L L^T \Lambda^{-1}  \nonumber\\
    &= (\Lambda^{-T} L) (\Lambda^{-T} L)^T 
\end{align}
and
\begin{align}
    B = \Lambda^{-T} L.
\end{align}
Since $\Lambda$ and $L$ are invertible, so is $B$.

To recap, one chooses a suitable $C$ and then constructs $D\rightarrow \Lambda\rightarrow M\rightarrow L\rightarrow B$. The details of finding these matrices can be found in Appendix \ref{sec:matrix_algorithms}. Once $B$ and $C$ are known, we construct $B^{-1} C B = A$, and the set $\{ A, A^2, A^3, ... A^{N-1} \}$ provides a perfect solution.

One will find that any solution created this way will contain the identity in the set $\{ A, A^2, A^3, ... A^{N-1} \}$. This seems to indicate that there is a canonical set of solutions. There are more solutions which are not of this form. To find some of these solutions, take any invertible matrix $\Delta$, and any perfect solution $\{ A, A^2, A^3, ... A^{N-1} \}$, and construct $\{ \Delta A \Delta^T, \Delta A^2 \Delta^T, \Delta A^3 \Delta^T, ... \Delta A^{N-1} \Delta^T \}$. These solutions are obviously symmetric. Their differences are also invertible:
\begin{align}
    (\Delta A^j \Delta^T - \Delta A^k \Delta^T)^{-1} 
    &= (\Delta (A^j - A^k) \Delta^T)^{-1} \nonumber\\
    &= \Delta^{-T} (A^j - A^k)^{-1} \Delta^{-1}.
\end{align}
More solutions can be found this way, but we will focus on canonical solutions.

\subsection{Family Lookup} \label{sec:lookup}

Practically speaking, we also need a lookup algorithm that takes a string as an input and returns the family that contains that string. Fortunately there is a fast algorithm which uses the permutation from Eq. (\ref{avvp}). Each family $k$ has a permutation associated with it defined by:
\begin{equation}
    P^{(k)}(i) = \underset{k\text{ times}}{P(P(P(...P(}i))))
\end{equation}
$P^{(N-1)}$ must be the identity permutation because $C^{N-1}$ is the identity, so these permutation form a group. Finding the family that the string $S_{i,j}$ belongs to is equivalent to solving:
\begin{equation}
    P^{(k)}(i) = j\label{eq: onetoone}
\end{equation}
for $k$. Since each string exists in a family, this equation has a solution for all $0<i,j<N$, making the set of permutations transitive. An efficient way to solve this is to define the permutation:
\begin{equation}
    Q(k) = P^{(k)}(1)
\end{equation}
We know that $Q(k)$ is one-to-one in the domain $0<k<N$, because the group is transitive, so it also has an inverse permutation $Q^{-1}(k)$. This can be calculated once quickly by applying the permutation $P$ on 1 repeatedly, and storing it in memory. (The choice of 1 is arbitrary.) Now our equation can be solved using $i = P^{Q^{-1}(i)}(1)$.
\begin{align}
    P^{(k)}(P^{(Q^{-1}(i))}(1)) = P^{{(Q^{-1}(j))}}(1) \\
    k + Q^{-1}(i) = Q^{-1}(j) \text{ mod } (N-1)
\end{align}
This is a modular equation which is trivial to solve since $Q^{-1}$ is known.

\subsection{Diagonalization}
\label{sec:diagonalization}

Once a suitable set of commuting Pauli strings is found, it is necessary to simultaneously diagonalize the family in order to measure the strings. There is a surprising method to do this with canonical solutions, where each family is diagonalized by the generating matrix of what is typically a \emph{different} family. Diagonalization can be done by finding a unitary transformation which maps the current family to the diagonal ($z$) family. In this section we describe how to construct these unitary transformations.

We begin by examining transformations of the form
\begin{align}
    U = \exp\left(\frac{i \pi}{4} x_k\right).
    \end{align}
Acting on the strings of a family,
\begin{align}
    U z_i x_{P(i)} U^\dagger = \left\{ \begin{array}{cc}
        z_i x_{P(i)} & \text{if } f(i,k) = 0 \\
        -i z_i x_{P(i) \oplus k} & \text{if } f(i,k) = 1 
    \end{array} \right\} \nonumber \\
    = (-i)^{\bar f(i,k)}z_i x_{P(i) \oplus (k \cdot \bar f(i,k))} .
\end{align}
Here $\bar f$ is equal to a cast of $f$ into the real numbers in the obvious way. Now consider a more general transformation with some set $K$ of $x$-strings:
\begin{align}
    U = \exp\left( \frac{i \pi}{4} \sum_{k \in K 
    } x_k\right) \\
    U z_i x_{P(i)} U^\dagger= \left(\prod_{k\in K} (-i)^{\bar f(i,k)}\right) z_i x_{P(i) \oplus \sum_k k \cdot \bar f(i,k)}.  \label{UzxU}
    \end{align}
 In order to reach the $z$ family we must have 
\begin{align}
    P(i) \oplus \sum_{k\in K } k \cdot \bar f(i,k) = 0.
    \label{diagcondition}
\end{align}
This result gives a condition for the diagonalization. The phase in (\ref{UzxU}) will need to be calculated later. First let us rewrite the diagonalization condition (\ref{diagcondition}) using the binary vectors (recall that $\oplus$ refers to converting integers into to $\mathbb{Z}_2^m$ vectors and then performing addition mod 2). Then (\ref{diagcondition}) is equivalent to:
\begin{align}
    \sum_{k\in K } v_k \Big( \sum_b (v_k)_b (v_i)_b \Big)= v_{P(i)}
\end{align}
or rearranging and expanding in components,    
\begin{align}
    \sum_b \Big(\sum_{k\in K}  (v_k)_a (v_k)_b \Big)(v_i)_b = \left(v_{P(i)}\right)_a.
    \label{diagcondition2}
\end{align}
Comparing with Eq.~(\ref{avvp}), we see that (\ref{diagcondition2}) will be solved if the set $K$ is chosen such that
\begin{align}
    \sum_{k\in K}  (v_k)_a (v_k)_b = A_{ab}.
\end{align}

To find the set $K$ and the relevant $v_k$ we proceed as follows. Let $K$ be of order $m$ and enumerate its elements as $k_a$, $a=1\dots m$. Then
 define a matrix $B_{ab} = (v_{k_a})_b$, such that $A = \sum_i B_{ia} B_{ib}$. If $B$ is symmetric then $A = B^2$. Since the powers of $A$ are cyclic due to the properties of companion matrices, there is always a symmetric matrix $A^{N/2}$ which is equal to $B$:
\begin{equation}
    A^{N/2}_{ab} = B_{ab} = (v_{k_a})_b.
\end{equation}

In this way, the $x$ strings needed to perform the diagonalizing transformation can be extracted from the generating matrix. In general, we can diagonalize the family defined by $A^i$ by using $(A^i)^{N/2}$. This is an efficient way to diagonalize the family which can be faster than performing a brute force diagonalization algorithm. 
$(A^i)^{N/2}$ can be found via:
\begin{equation}
    (A^i)^{N/2} = \left\{ \begin{array}{cc}
        A^{i/2} & \text{if } i \text{ mod } 2 = 0 \\
        A^{\frac{N+i-1}{2}} & \text{if } i \text{ mod } 2 = 1
    \end{array} \right\}.
    \label{AiN2}
\end{equation}
In fact the matrices on the right-hand side of Eq.~(\ref{AiN2}) are calculated already when generating the families, so the most efficient method is to construct the measurement bases at the same time as organizing the families. In Appendix \ref{sec:circuit_rep} we review the map from the diagonalizing unitaries thus constructed to  quantum circuits.

There are $m$ rows of these matrices, so it takes $m$ transformations to get from a family in a perfect solution to the $z$ family. This is the minimum number of transformations to take a family to a new family with no shared strings, because any string not in a commuting subgroup commutes with exactly half of the strings in that subgroup minus 1, since the subgroup is closed under multiplication, and doesn't include the identity. Appendix \ref{sec:circuit_rep} shows that each of these transformations can be written in $2m + 1$ gates, so the total circuit depth for the entire family should scale as $m^2$.

To calculate the phase in (\ref{UzxU}), recall that if $f(i,j) = 1$, then $e^{(i \frac{\pi}{4} x_i)} P_j e^{(-i \frac{\pi}{4} x_i)} = P_j (-ix_i)$, which means that each transformation gives a right multiplication of $-ix_i$. When diagonalizing strings, a series of transformations is used, so there will be a right multiplication of  $ -ix_k$ for all strings $x_{k\in K}$  which don't commute with the $P_j$ being diagonalized. To formalize this, for each $P_j$ in a family there will be a subset $H \subseteq K$ such that $\forall k \in H: [x_k, P_j] \neq 0$. It is this set that transforms $P_j$ and contributes to the phase. Suppose $P_j$ has the decomposition $P_j = e^{i \theta} z_a x_b$. The phase here is a power of $i$ that arises because of the fundamental relation $\sigma_y = -i \sigma_z \sigma_x$.  (The accumulated phase in a string from this decomposition was mentioned previously in the discussion below Table~\ref{tab:product table} but is neglected in the table; we must now account for it.) In other words, every time a $Y$ appears in a string, there is an additional factor of $-i$ in the decomposition. We  define $Y(a,b)$ as a function which returns the number of $Y$s in $P_j$.
Then:
\begin{equation}
    U P_j U^{\dagger} =  (-i)^{Y(a,b)} z_a x_b \Big( \prod_{k \in H} -i x_k\Big) 
\end{equation}
The product of the $x$ strings must be an $x_b$, so the product in each tensor space has the form:
\begin{align}
    U P_j U^{\dagger} = (-i)^{N_H + Y(a,b)}z_a
\end{align}
where $N_H$ is  the cardinality of $H$. 
Thus the net phase factor multiplying the resultant $z$ string receives contributions both from the decomposition of the original string and the rotation. Computing it requires counting the commuting $x$ strings in the transformation, and calculating the $zx$ decomposition.

\section{Implementation on IBM hardware using {\tt qiskit}} 
\label{sec: Implementation}
\begin{figure}[t!]
\centering
\includegraphics[width=0.45\textwidth]{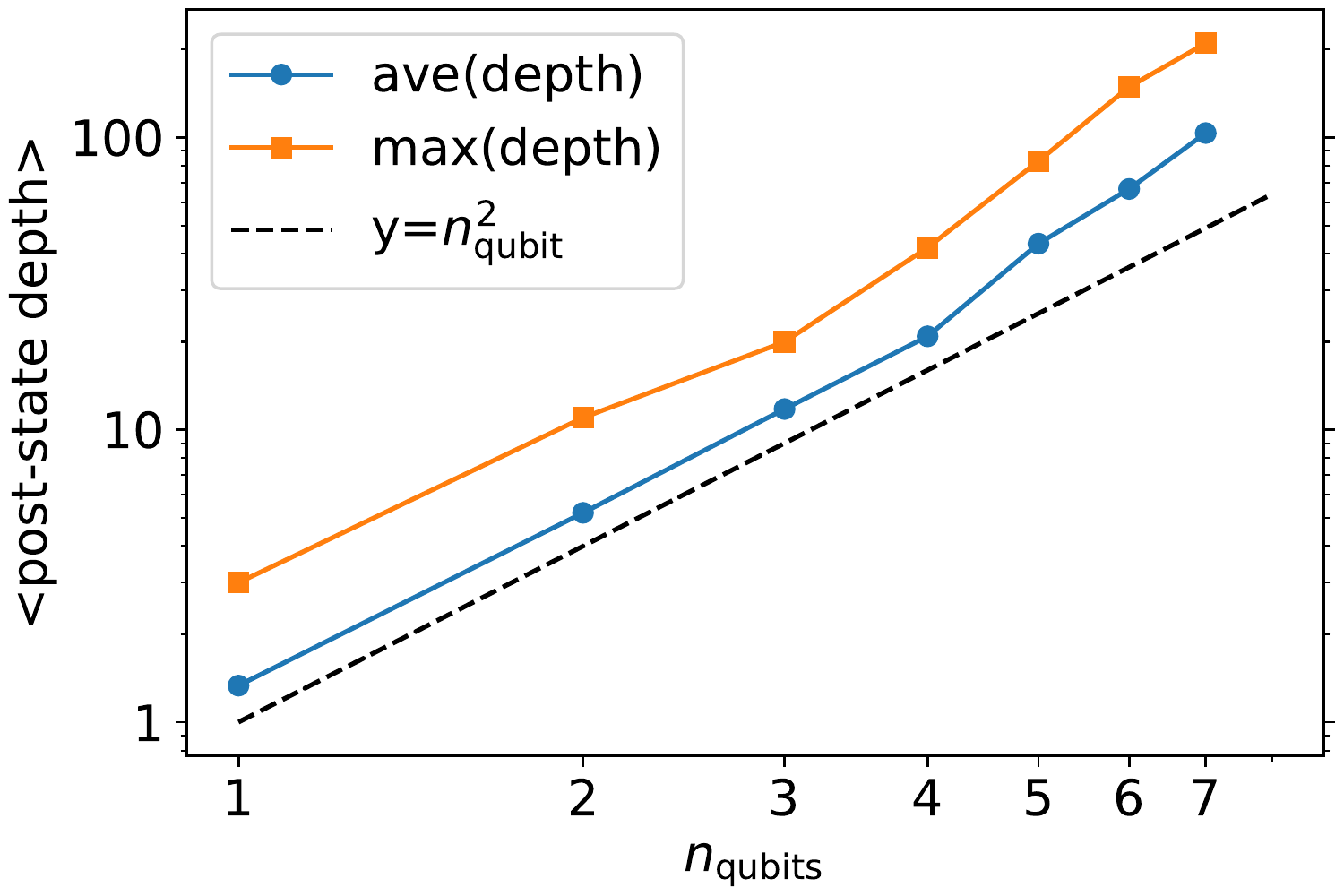}
\caption{
Post-state circuit depths after transpilation, targeting the quantum hardware {\tt ibm\_oslo}. The orange line shows the increase in
circuit depth for the family whose post-state circuit rotation is deepest. The blue line gives depth increase averaged over all $2^m+1$ families.
The dotted line shows that the depth increase scales with qubit count $n_{\rm qubits}=m$ approximately as $m^2$.
}
\label{fig:poststate-depth}
\end{figure}

In principle, for fully dense Hamiltonians where all Pauli strings contribute, our dense grouping algorithm will give a quadratic improvement over naive evaluation of each string individually.  Because the dense grouping method is optimal in the sense that it generates the fewest number of families which can partition all $4^m$ strings, it should also improve upon other available methods when applied to sufficiently dense, though not necessarily fully dense, Hamiltonians or other observables of interest. Such problems do arise in interesting physical models, and we consider an application to VQE on a particular example of such a model in Sec.~\ref{sec:vqe}.

For real applications, there are different metrics relevant for assessing the cost-benefit of dense grouping as compared to other methods. These include the number of shots required for achieving a target precision, the time required for classical simulation, runtimes on quantum hardware, and accuracy for fixed resource use.

To study these, we have integrated our Python package that generates the Pauli groupings into IBM's {\tt Qiskit} framework, which can be used to run quantum circuits on hardware from IBM and other vendors, and may also be used for classical simulation of quantum devices.
Both the Python package and {\tt Qiskit} integration we have made open source. More information on implementation details and usage of these are given in Sec.~\ref{sec:code}.

As a benchmark, in addition to naive evaluation, in what follows we will compare results using the dense grouping with the native {\tt Qiskit}~\cite{Qiskit} grouping of Pauli strings into families, based on a greedy graph coloring algorithm over qubit-wise commuting (QWC) Pauli strings. This functionality is provided by the {\tt AbelianGrouper()} class,
and for a fully dense Hamiltonian (all $4^m$ Pauli strings have non-zero coefficients) results in a partition into $3^m$ families.\footnote{The $3^m$ scaling can be seen as follows. Consider the set of Pauli strings which do not contain $I$ in their string representation. None of these strings qubit-wise commute with each other, so any QWC set of families will necessarily need a family for each of these strings, for a total of at least $3^m$ families. Then, any strings which do contain $I$s  in their string representation qubit-wise commute with the corresponding string where all of the $I$s are replaced by $Z$s. These strings have already been sorted into families, so no additional families are needed, establishing the claim. }

For the dense grouping strategy, the elements of the families are generally commuting (GC), except for the $x$, $y$, and $z$ families which are QWC. The circuits required to transform a given family to the $z$ family are described in more detail in Appendix~\ref{sec:circuit_rep}. We call these operations post-state rotations, because they are applied at the end of the circuit which generates $| \psi \rangle$ for the expectation value of interest. Circuits for the evaluation of GC families will have increased depth as compared to QWC families, due to the non-trivial post-state rotations required to bring them to the $z$ family. The average and maximal post-state circuit depth as a function of $n_{\text{qubits}}$ is shown in Fig.~\ref{fig:poststate-depth}. (In this and subsequent sections, for ease of reading figures, we will use $m$ and $n_{\text{qubits}}$ interchangeably.)  In contrast, circuits grouped into QWC families have post-state rotation depth of 1.

As a practical matter, the increased circuit depth can impact the efficacy of our method relative to QWC strategies. Circuits for GC families will have increased runtime which can modify scaling arguments based only on the number of families, and for noisy devices increased depth may affect the precision of results. We find however, as discussed in more detail in the following two sections, that the effects of increased state depth are relatively insignificant for the ranges of $n_{\text{qubit}}$ studied here, so that the algorithmic improvement of dense grouping is close to that of scaling arguments 
based on the number of families.

\subsection{Computational cost}

Here we compare the integrated quantum circuit runtime for evaluating expectation values using the Abelian and dense methods, allocating a fixed number of shots per group/circuit. As a simple model for the runtime of a single circuit, we take
\begin{equation} \label{tau_def}
    \tau = \tau_{\text{over}} + \tau_{\text{circ}}(d) \,;
\end{equation}
in $\tau_{\text{over}}$ we include all circuit overhead costs except those for running the circuit, $\tau_{\text{over}} = \tau_{\text{reset}} + \tau_{\text{delay}} + \tau_{\text{meas}}$~\cite{Tornow_2022}, and $\tau_{\text{circ}}(d)$ is the time to run a circuit of depth $d$. We assume the runtime is approximately linear in the transpiled state depth $d$. 
Let the state preparation circuit depth be $D$.
We write the average post-state depth over families as $\text{ave}_f[d_{\text{post}}(f)]$ and
define $a=\text{ave}_f[d_{\text{post}}(f)]/m^2$, to make the scaling of post-state depth with $m$ explicit. $a$ itself will depend weakly on $m$, and we tabulate values of $a$ obtained using the {\tt ibmq\_quito} transpiler in Table~\ref{tab:avals}.
The ratio of timings for Abelian and dense methods can then be modeled as
\begin{equation} \label{tau_ratio}
\frac{\tau_{\text{Abelian}}}{\tau_{\text{dense}}} = 
\frac{3^m \Bigl[\tau_{\text{over}}+\tau_{\text{circ}}(D+1)\Bigr]}{(2^m+1) \Bigl[ \tau_{\text{over}}+\tau_{\text{circ}}(D+am^2) \Bigr]} \,.
\end{equation}
If the circuit overhead is much greater than the circuit runtimes $\tau_{\text{circ}}$, or if $D >> am^2$, the runtime improvement should be close to $(3/2)^m$. 
The same considerations hold for $\tau_{\text{naive}}$ with $3^m \to 4^m$ in the numerator of~\eqref{tau_ratio}.
If $am^2 \gtrsim D$, the accuracy of the expectation value could be impacted for NISQ hardware or noisy simulators, relative to a QWC strategy.

\begin{table}[h]
    \centering
    \begin{tabular}{c|c|c}
         m & a (transpiled) & a (nominal)\\
         \hline
        2 & 1.30 & 0.8\\
        3 & 1.23 & 0.75 \\
        4 & 1.23 & 0.77 \\
        5 & 1.54 & 0.77
    \end{tabular}
    \caption{Empirical $a$ values, measured from post-state rotation circuits transpiled with the {\tt ibmq\_quito} backend. The third column (nominal) gives values based on the circuit description
    in the text.}
    \label{tab:avals}
\end{table}

\subsection{Runtimes on IBM hardware} \label{sec:runtimes}
We ran a series of tests on {\tt ibmq\_quito} to compare
runtimes using the Abelian and dense groupings. Device characteristics measured near the time of running are collected in Appendix~\ref{sec:device}.
We prepared the state using the {\tt EfficientSU2} ansatz circuit, with randomly assigned phases for the parameters (these were held fixed between runs).
In order to distinguish the different terms in
Eq.~\eqref{tau_ratio}, we varied the depth
of the state by increasing the {\tt reps} argument to
the ansatz function, which appends multiple repetitions
of the specified state circuit. We took {\tt reps} from 1 to 5. Timings were all obtained using 20k shots.

\begin{figure}
    \centering
    \includegraphics[width = 0.48\textwidth]{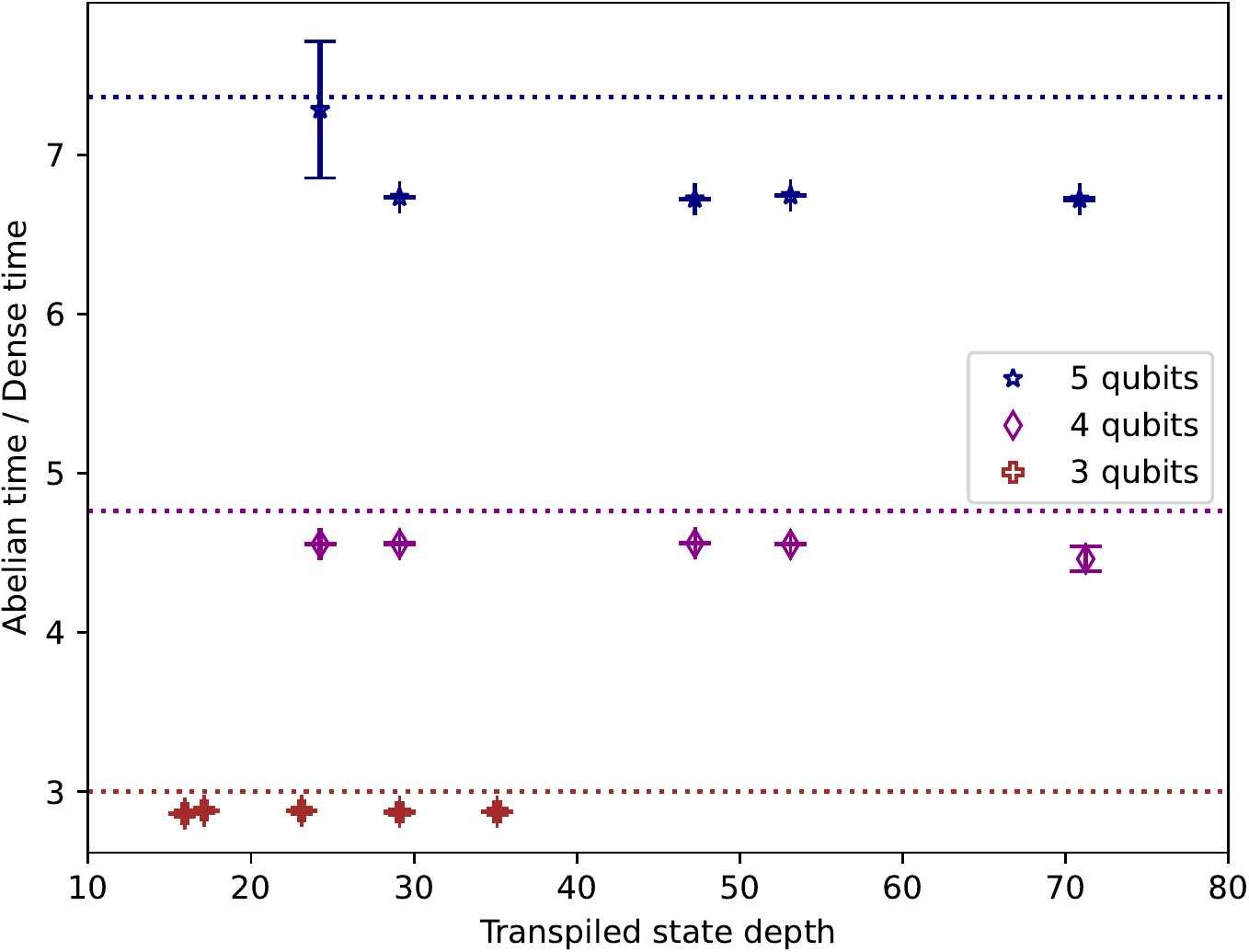}
    \caption{Circuits used to compute expectation values were calculated on {\tt ibmq\_quito} using both dense and Abelian grouping methods for 3 to 5 qubits. The ratios of the calculation times between different methods are shown here. The ideal speedup factor is the ratio of the number of circuits, $\frac{3^m}{2^m + 1}$ shown by dotted lines. The states measured are constructed using {\tt EfficientSU2}, and the {\tt reps} parameter is varied from 1 to 5 to show the effects of increased circuit depth. Note that in two runs the time taken was much larger than expected, leading to large error bars (determined by the standard deviation in a sample of three to five runs) around (25,7) and (70,4.5). It is unclear the cause of this error.}
    \label{fig:timing}
\end{figure}

The results in Fig.~\ref{fig:timing} show that increased state depth has a small effect on the time to complete calculating the expectation value, of the order of a few percent. Instead, the circuit overhead appears to be the largest contributing factor. As a result, the improvements obtained using the dense grouping are near to the ideal result based on counting families. Note that time required for mitigation runs, common to both methods, are not included in these timings.

At present, it is typical for quantum resources to be priced on a per-shot basis---These costs range from \$0.0002 to \$0.01 per shot depending on vendor and device~\cite{amazonQuantumComputer}. 
With a fixed number of shots allocated per group/circuit, the improvement provided by dense grouping is the ratio of the numbers of groups, provided accuracy is comparable between different methods (ignoring any fixed overhead independent of method if mitigation circuits are used).  
We study this in the next section and show that expectation value results for the dense grouping compare well with naive and Abelian methods with a fixed number of shots per group.

\subsection{EV accuracy}
 In general, one expects that the grouping method can impact precision due to the correlation of shots data across multiple operators in a family~\cite{Gokhale2019MinimizingSP}. For the naive grouping method, all Pauli strings are evaluated separately, whereas for grouped strategies the same shots data is used to evaluate contributions of all the operators in the group, which introduces correlations into the expectation value. The structure of these correlations will depend on both the grouping method and detailed form of the Hamiltonian and state, however it was shown in~\cite{Gokhale2019MinimizingSP} that without any prior on the state $|\psi \rangle$ the expected value of the covariance between any two commuting Pauli strings is zero.

In addition to covariance effects introduced by grouping, there is another effect specific to groupings containing generally commuting (as opposed to qubit-wise commuting) strings. For families of QWC strings, the post-state rotation circuit  to bring the family into the $z$ family will have depth 1. For GC families, the post-state circuit depth will depend on family and will be greater than 1. For the optimal dense groupings generated in this work, this depth on average grows with qubit number like $m^2$, as discussed in Appendix~\ref{sec:circuit_rep} and shown in Fig.~\ref{fig:poststate-depth}. 

Here we investigate the accuracy of expectation values
obtained using the dense grouping method and compare these with the naive and Abelian methods on the IBM simulator {\tt FakeOslo}~\cite{FakeOslo}.
We present results for the $A_1^+(g)$ Hamiltonian, discussed in more detail in the following section. We have also carried out analogous studies with random dense Hamiltonians and find similar results, but working with a physical Hamiltonian has the advantage that results can be directly compared while varying $n_{\text{qubit}}=m$. 

\begin{figure}
    \centering
    \includegraphics[width=0.49\textwidth]{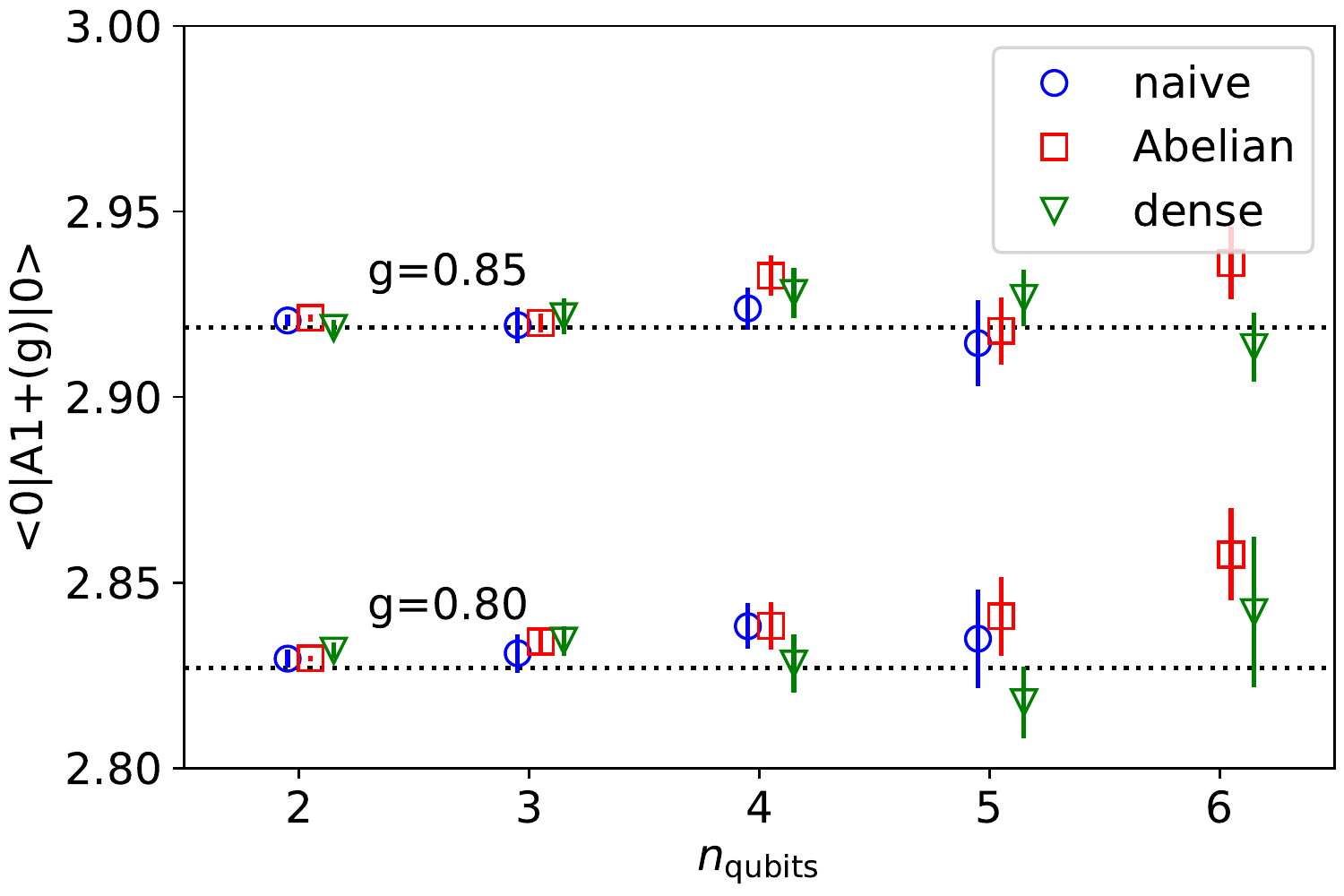}
    \caption{Expectation value measurements of the $A_1^+(g)$ Hamiltonian, as a function of $n_{\text{qubits}}$ and for different grouping strategies.
    The exact values of $\langle 0| A_1^+(g)|0\rangle$ for $g=0.8$ and $g=0.85$ are given by dashed lines indicated on the figure. Results obtained using the {\tt FakeOslo} simulator are shown for naive (blue circles), Abelian (red squares), and dense (green triangles) grouping strategies. The values shown here are averaged over 20 runs, with 20,000 shots on each run and using {\tt CompleteMeasFitter} error mitigation with 10,000 shots.}
    \label{fig:A1plus_EV_FakeOSlo}
\end{figure}

We carry out repeated evaluations of the expectation value of the Hamiltonian in the zero state, which we denote $\langle 0| A_1^+(g) |0 \rangle$. Note that here the zero state refers to the product state of all qubits in the zero state, rather than the ground state of the Hamiltonian, which is in general much more complicated because the Hamiltonian is dense. Working with the zero state maximizes any impact of post-state depth with respect to the naive and Abelian methods (i.e.\ to make the comparison conservative). We find that in the absence of error mitigation techniques, biases can be present that skew the expectation values for all grouping methods, and moreover that in each case these biases are effectively eliminated by applying error mitigation. The results shown in Fig.~\ref{fig:A1plus_EV_FakeOSlo} use the {\tt Qiskit} function {\tt CompleteMeasFitter} with 10k shots to compute the mitigation matrices. {\tt CompleteMeasFitter}~\cite{completemeasfitter} runs circuits over the full set of computational basis states to construct a calibration matrix, $M$,
and then minimizes $|C_{\text{noisy}} - M C_{\text{mitigated}}|^2$ with the observed result $C_{\text{noisy}}$ to find the mitigated result $C_{\text{mitigated}}$.

We compute $\langle0| A_1^+(g)|0 \rangle$ for coupling values $g=0.8, 0.85$ and $m \in [2, 6]$ using 20k shots/circuit for each of naive, Abelian, and dense groupings (we omit naive data for $m=6$ because the computation time became excessive on a laptop). In each case the computation is repeated 20 times, and the average and standard deviation of the results are plotted in Fig.~\ref{fig:A1plus_EV_FakeOSlo}. We find that all grouping methods are able to accurately predict the true expectation value in this state within errors, with the occasional $\approx 1 \sigma$ fluctuation. The uncertainties increase with the number of qubits $m$, ranging from $\lesssim 0.1 \% $ at $m=2$ to $\approx 0.5 \% $ at $m=6$. The uncertainties are commensurate across the three methods, with no method obviously outperforming another. This indicates that the effect of additional post-state depth incurred by the dense grouping is mild up to $m=6$, and that any of the methods can be used (with mitigation) to obtain accurate estimates. Note that the computational expense is least for the dense grouping; if computational expense was fixed the dense (and Abelian) uncertainties would be reduced relative to the naive grouping.

\section{Testing with VQE}
\label{sec:vqe}
The variational quantum eigensolver (VQE) algorithm relies on repeated evaluations of expectation values of a Hamiltonian $H$ with a given parameterized ansatz state to determine the minimum energy expectation and corresponding state over the ansatz manifold. For dense Hamiltonians an optimal grouping into families may provide significant cost/resource benefits.

\subsection{Femtouniverse}
Here we investigate this using
a matrix quantum mechanics model obtained by the dimensional reduction of 4D SU(2) gauge theory on a spatial
3-torus, the so-called ``femtouniverse'' model~\cite{Bjorken:1979hv,thooft,Luscher:1982ma,LUSCHER1984445,VANBAAL1986548,van1988gauge,vanBaal1990,vBK,van1987qcd}. This model is interesting for isolating some of the nonperturbative low-energy dynamics of a confining gauge theory into a solvable quantum mechanics model, and because matrix models are thought to play a significant role in holographic theories of quantum gravity (albeit at large $N$.) The theory is most compactly formulated on a gauge-invariant state space, but constrained to such a space the Hamiltonian is generally dense. We have recently studied quantum simulations of the femtouniverse~\cite{Butt:2022xyn}, and here we exhibit the impact of adding Pauli grouping to the VQE determination of the ground state.

\begin{figure*}[t!]
    \centering
    \includegraphics[width = 0.45\linewidth]{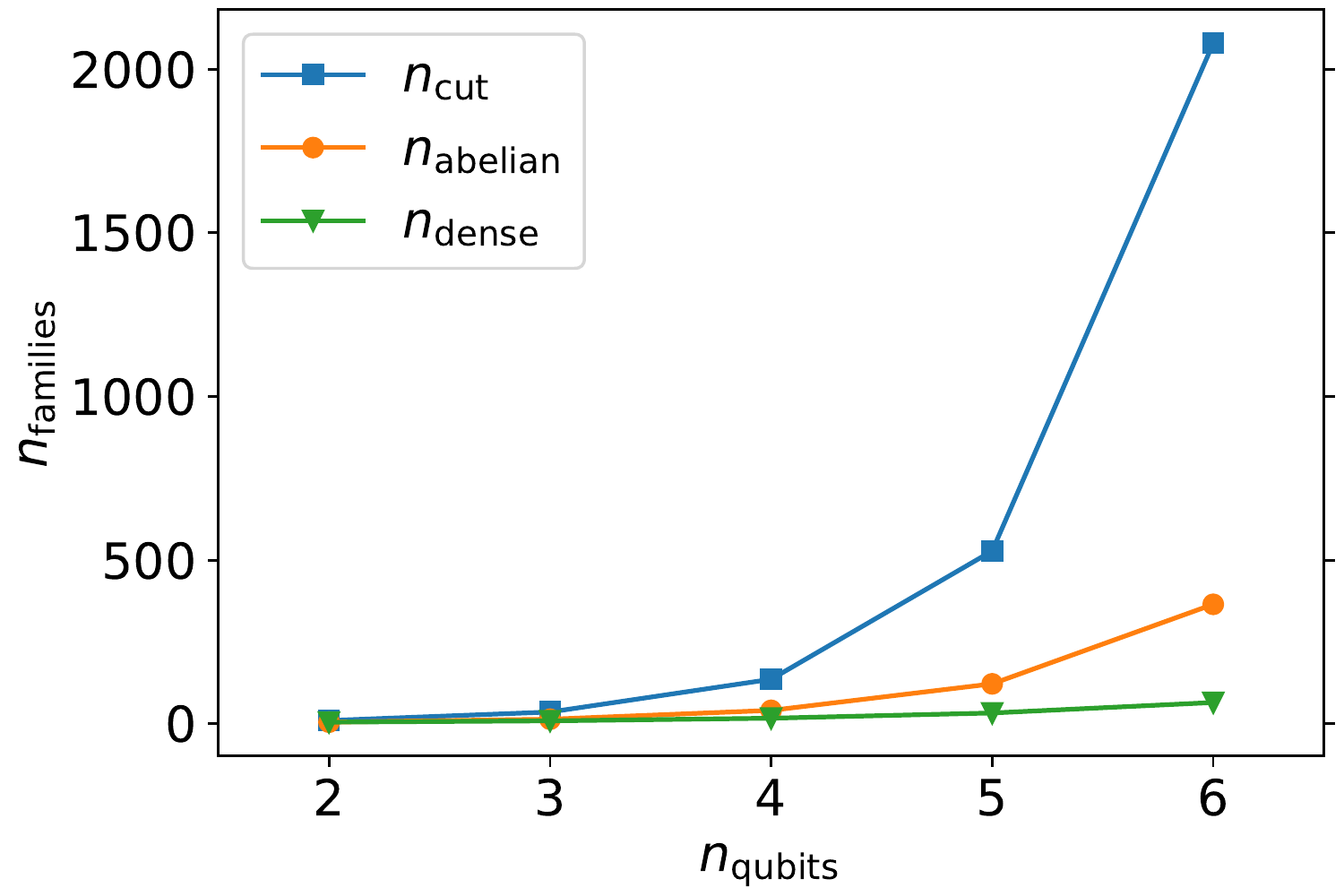}~~
    \includegraphics[width = 0.45\linewidth]{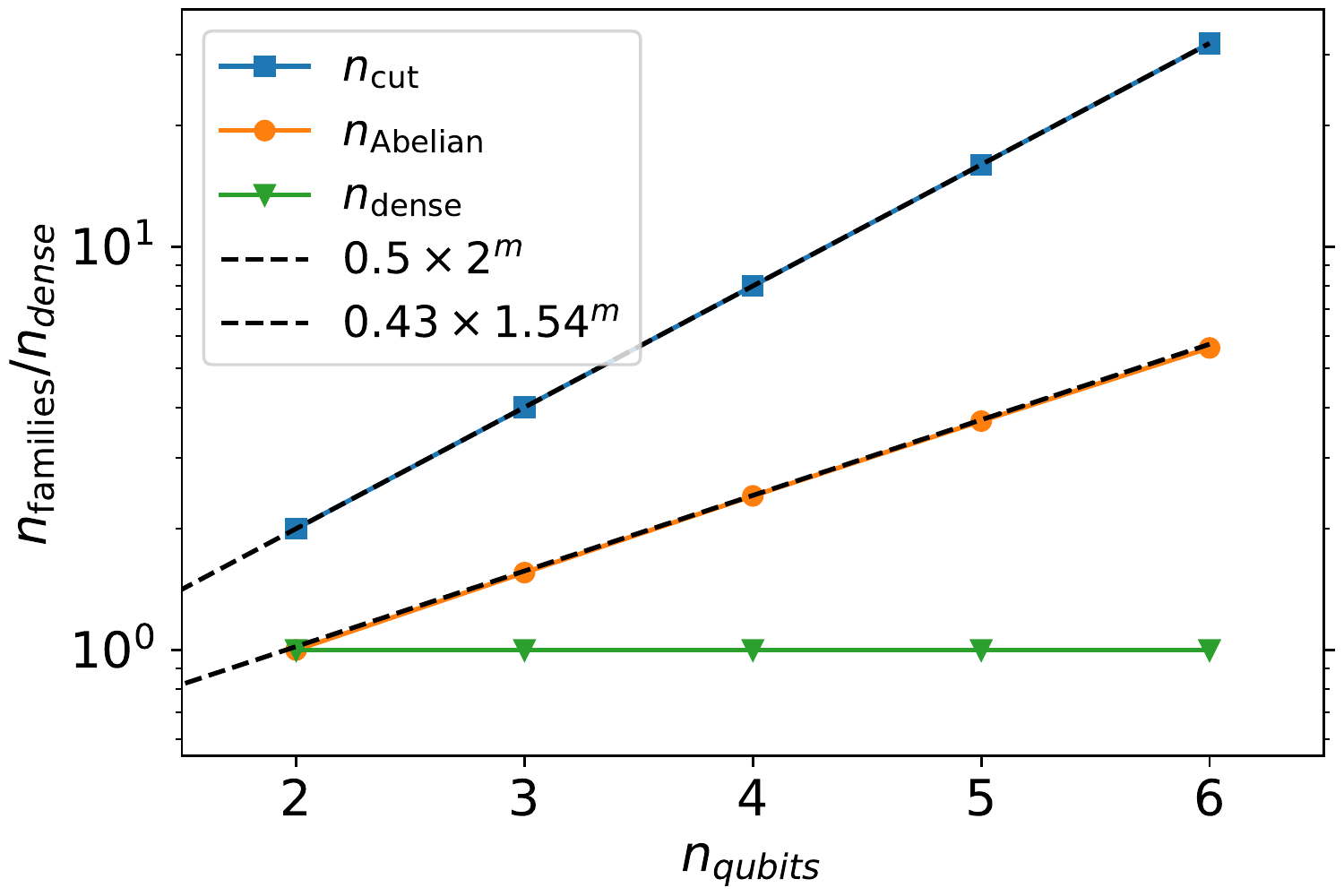}
    \caption{(Left) Number of family groupings generated by different grouping methods for the $A_1^+(g=0.8)$ femtouniverse Hamiltonian, as a function of number of qubits $m$. We compare the naive decomposition into individual Pauli strings, the {\tt AbelianGrouper}, and the dense grouping. (Right) The same data but plotted as a ratio to the number of families from the dense grouping ($2^m+1$),
    showing the improvement factor of the dense grouping compared to measuring individual Pauli strings (blue squares) and grouping generated by {\tt AbelianGrouper} (orange circles). The dotted lines give an indication of the exponential improvement observed using the dense vs.\ other methods. }
    \label{fig:nfamilies_A1plus}
\end{figure*}

\begin{figure}
    \centering
     \includegraphics[width=0.45\textwidth]{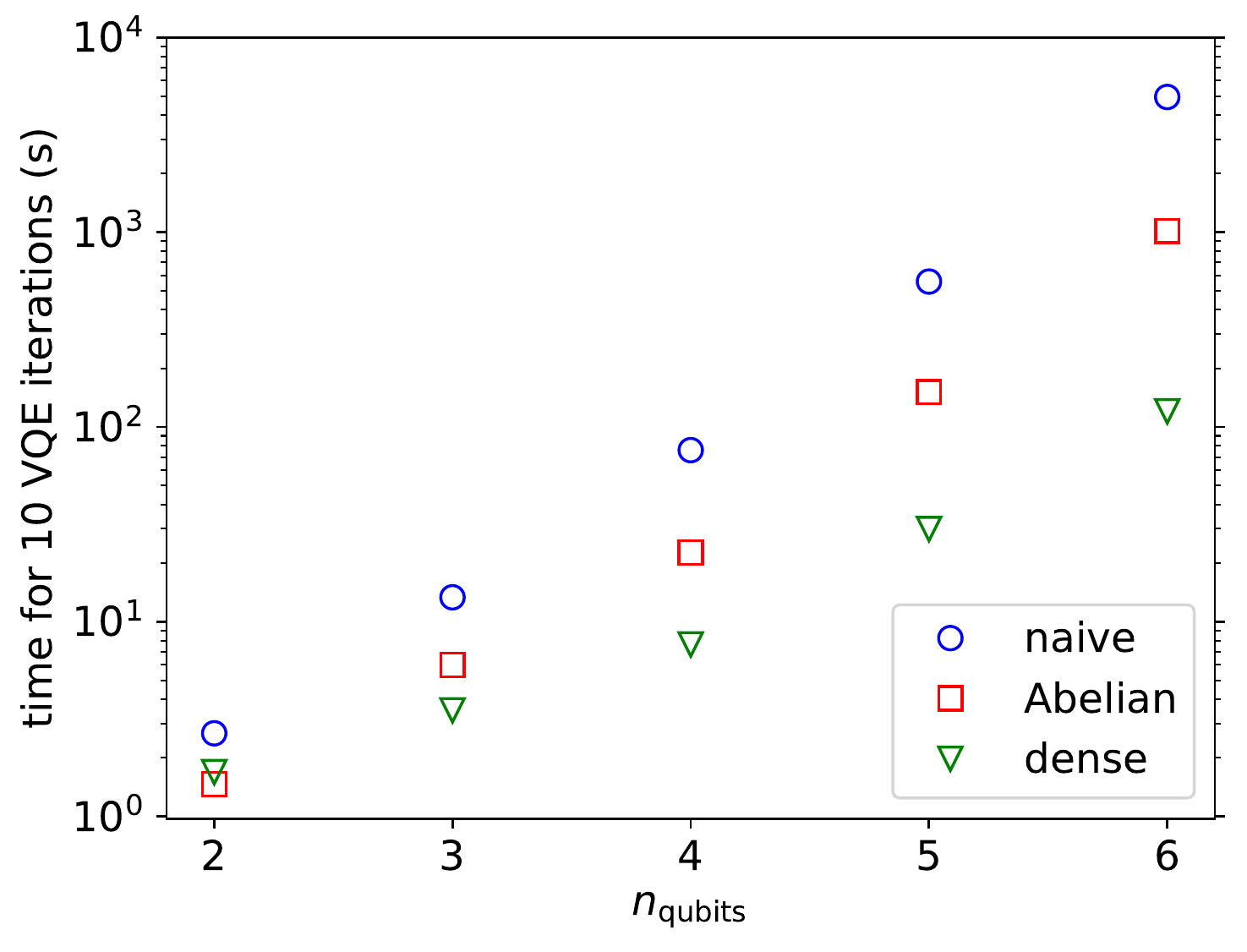}
    \caption{Time for 10 VQE iterations vs $n_{\text{qubits}}$ on AER simulator}
    \label{fig:vqe-time}
\end{figure}

We construct a truncated Hamiltonian in the zero-flux sector (known as the $A_1^+$ sector, for historical reasons). We decompose the Hamiltonian into Pauli strings and apply a cut to remove strings which have coefficients with absolute values less than a given tolerance. The numerical value of the cut was taken to be 0.0001 in units of the inverse spatial torus size $1/L$. We observe that the number of Pauli strings remaining after the cut is $\gtrsim 50\%$ compared to the fully dense scenario ($4^m$) for the range studied $m \in [2,6]$, and that the number of strings scales exponentially with $m$. We then group the resulting Hamiltonian using the dense partitioning, and we compare this with the grouping generated by the {\tt AbelianGrouper} class native to {\tt Qiskit}. The results of this are shown in Fig.~\ref{fig:nfamilies_A1plus}, where we plot the resultant number of families vs $m$.
For $m=2$, both the Abelian and dense groupings generate 5 families, for $m>2$ the dense grouping outperforms. At $m=6$ the dense grouping has an approximate $6 \times$ improvement with 365 vs.\ 65 families for the Abelian and dense groupings respectively, and based on observed scaling, for larger $m$ the improvement from the dense grouping will continue to increase. In Fig.~\ref{fig:vqe-time} we plot the time for 10 VQE iterations vs $n_{qubits}$ and observe an improvement consistent with the gain predicted by ratio between number of families from Abelian grouping vs. dense grouping. Where the dense grouping makes VQE computation tractable for higher number of qubits, it doesn't guarantee an improvement in the accuracy of VQE results since they depend upon multiple simulation parameters. 

\section{Description of Public Code}
\label{sec:code}

Our public code is presented in two packages. The first package is called {\tt psfam.py}, and its purpose is to partition strings, and to construct a unitary operator of clifford gates which diagonalizes each family. The second package is an extension of {\tt qiskit} that integrates the operator groupings provided by {\tt psfam} into expectation value measurements.

{\tt psfam} presents the partitioning for any number of qubits $m$. The following code prints the generating matrix $A$ from Eq.~(\ref{avvp}) and reports each family:
\begin{python}
from psfam.pauli_organizer import *
m=2
PO = PauliOrganizer(m)
print(PO.properties())
print()
for f in PO.get_families():
    print(f)
\end{python}
The output is:
\begin{python}
Qubits: 2
Generating Matrix:
[1, 1]
[1, 0]

XZ,YX,ZY
XY,ZX,YZ
IY,YI,YY
IX,XI,XX
II,IZ,ZI,ZZ
\end{python}

Additionally, this object has functionality to construct the post-state rotation circuits described in Sec.~\ref{sec:diagonalization} and Appendix~\ref{sec:circuit_rep} through the {\tt PauliOrganizer.apply\_to\_circuit(circuit)} method. It also calculates the coefficients from the end of Appendix~\ref{sec:circuit_rep} which represent the contribution of each measurement to the expectation value. This package can be found at \cite{psfam}.

The {Qiskit} extension {\tt dense\_ev}~\cite{denseev} provides two classes  used to compute expectation values using
the optimal dense groupings provided by {\tt psfam}.
The first of these is {\tt DenseGrouper}, which
converts a {\tt SummedOp} object into a sum of {\tt SummedOp} objects (also of type {\tt SummedOp}), organized according to the specification of {\tt psfam}.

The second class, {\tt DensePauliExpectation}, contains the logic to compute expectation values on quantum hardware or simulators using optimal dense grouping.
It can be used as a replacement for the Qiskit native
{\tt PauliExpectation} class. The code fragments below give examples of this for a simple expectation value,

\begin{pythonLines}
from dense_ev import DensePauliExpectation

# EV
...
    
ev_spec = StateFn(H).compose(psi)
expectation = \
DensePauliExpectation().convert(ev_spec)

...

\end{pythonLines}
and as a component of the VQE algorithm:
\begin{pythonLines}
# VQE
...

vqe = VQE(ansatz, optimizer=spsa, 
    quantum_instance=qi,
    callback=store_intermediate_result,      
    expectation=DensePauliExpectation())
    
result =\
vqe.compute_minimum_eigenvalue(operator=H)

...

\end{pythonLines}

We also provide routines that perform unit tests on our codes, and support for the Qiskit {\tt Estimator} primitive.

\section{Comparison with graph-theoretic methods}
\label{sec:graphcomparison}
\begin{figure*}
\centering
\includegraphics[width=0.49\textwidth]{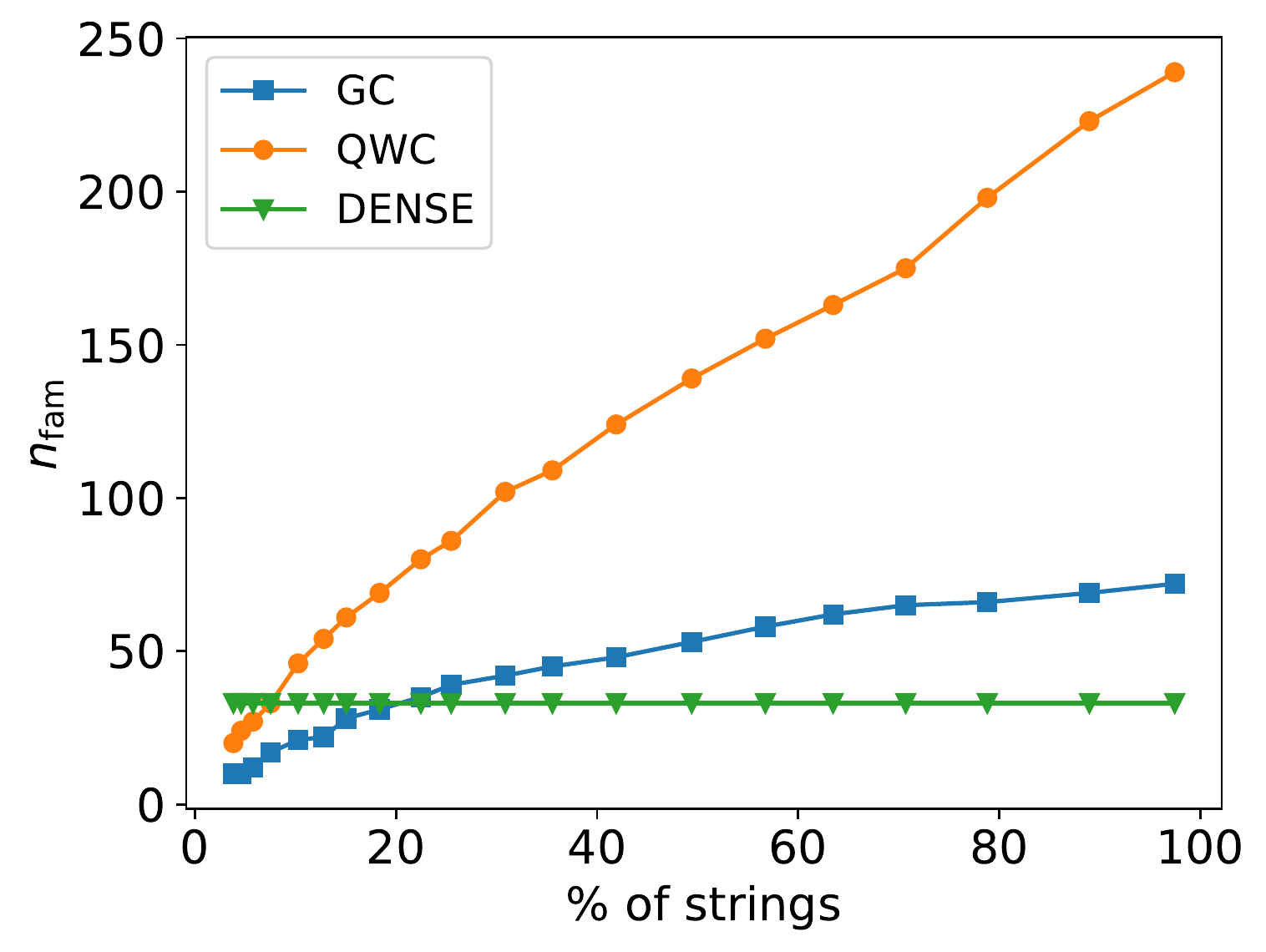}
\includegraphics[width=0.49\textwidth]{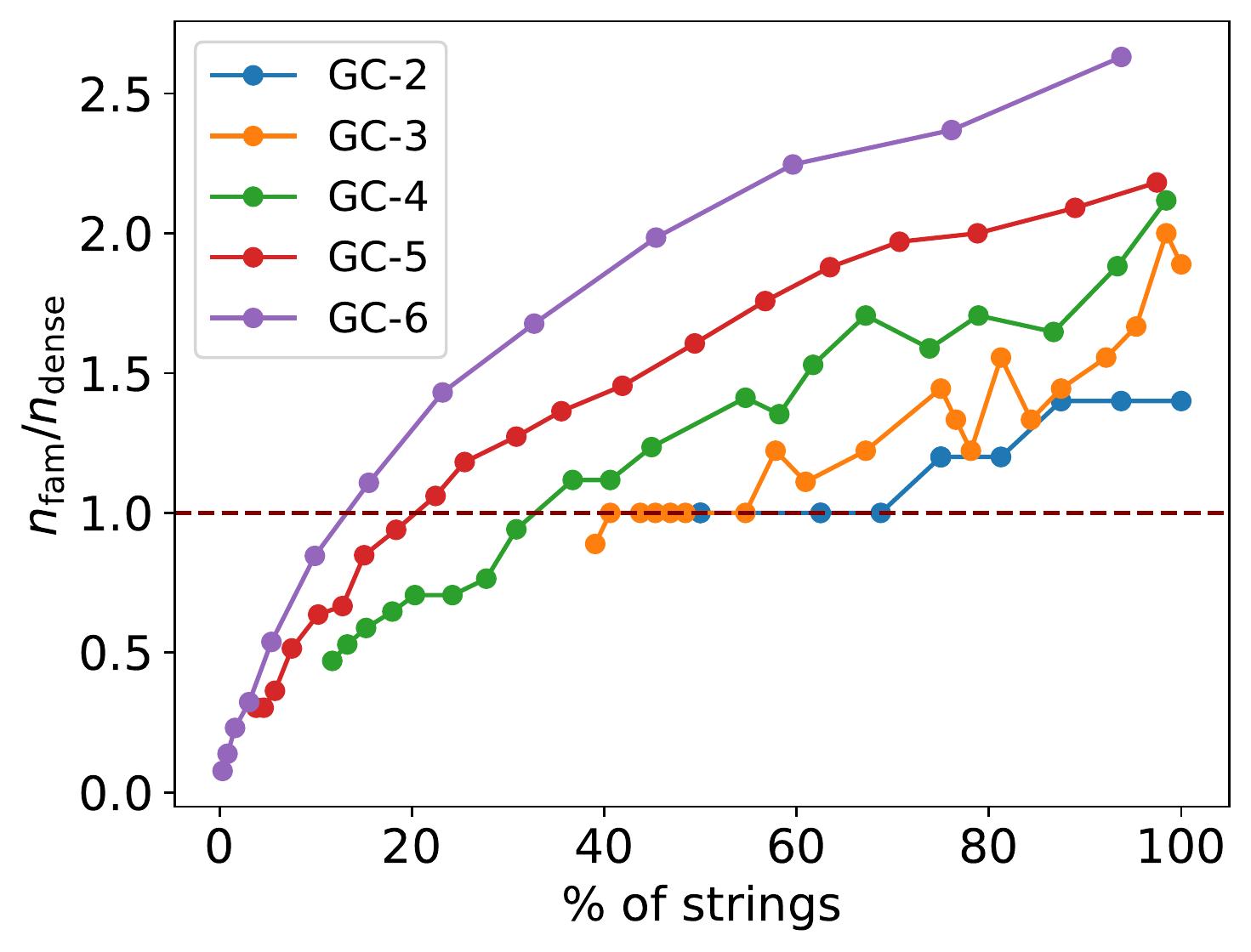}
\caption{(Left) Starting with a random (fully dense) Hermitian operator, with $m=5$,
a numerical cut of increasing magnitude is applied on the coefficients of the Pauli string decomposition to reduce the number of strings present. The resulting operator is grouped according to the GC, QWC, and DENSE methods. The x-axis gives the percentage of the $4^m$ original strings in the operator and y-axis gives the number of family groupings for each method.
(Right) A numerical cut of increasing magnitude is applied to a random Hermitian operator, for $m \in [2,6]$. The corresponding lines are labelled GC-2 (lowest curve at the right of the figure) through GC-6 (top curve). The y-axis now shows the ratio of the number of families produced by GC to that of DENSE. For visual clarity, a horizontal dotted line is plotted at 1, with points above/below the line indicating outperformance by DENSE/GC.}
\label{fig:nfam}
\end{figure*}

In this section we study in more detail the performance of DENSE grouping compared to graph theory-based methods. The problem of partitioning Pauli strings can be expressed as a graph theory problem~\cite{Gokhale2019MinimizingSP,Verteletskyi_2020,yen2020measuring,Jena:2019ddb,izmaylov2019unitary}, where strings are represented as nodes, and edges between nodes represent whether two strings commute (or anti-commute, depending on the exact problem formulation). 
We make comparisons with the Largest First algorithm made available through Qiskit, implemented in the {\tt rustworkx} package~\cite{Treinish2022}. 
We study the sizes of solutions generated via different methods,
and the classical computing resources required to generate those solutions.

For dense matrices  where the number of Pauli strings $N_{\text{Pauli}} > 4^m-2^m$, the DENSE algorithm provides a grouping with the minimal number of families, but in principle a graph algorithm could also find optimal or near-optimal solutions. As the density of the matrix considered (as measured by the number of different Pauli strings in its decomposition) is reduced, because the DENSE solution is essentially fixed to $2^m +1$ families (unless some families happen to be empty), we expect that at some point graph algorithms will generate a solution with fewer family groupings. We investigate this using random (fully dense) Hermitian matrices, which are made less dense by applying a numerical cut of increasing magnitude on the Pauli string coefficients. 

Fig.~\ref{fig:nfam} (left) shows an example of this for a matrix with $m=5$. In this example, the DENSE method outperforms until the number of Pauli strings is roughly $20 \%$ of the original $4^m$ strings in the fully dense matrix. We checked that this behavior is ``typical'' for these matrices by observing roughly consistent behavior when changing the random seed used to generate the matrices. Fig.~\ref{fig:nfam} (right) illustrates the relative performance of GC vs.\ DENSE, as $m$ is varied. Again we find there is a range where dense, but not fully dense, matrices are grouped more efficiently using the DENSE algorithm.
Note that the performance of GC can depend on the details of the operator structure, i.e.\ the results we found by pruning random matrices may not hold for other classes of matrices. For example, we found that applying GC to the Femtouniverse Hamiltonian, which has a population $\sim 50 \%$, results in $2^m$ families, outperforming DENSE by one family.

We now turn to a discussion of the classical computing resources required to generate solutions via graph methods and DENSE. For fully dense operators, the walltime and memory resources needed for both methods scale exponentially in $m$, but as we will see the DENSE method has a roughly square root improvement over the graph methods, so that for fixed computational resources the $m$'s that can be practically reached are significantly larger.

\begin{figure*}
\centering
\includegraphics[width=0.49\textwidth]{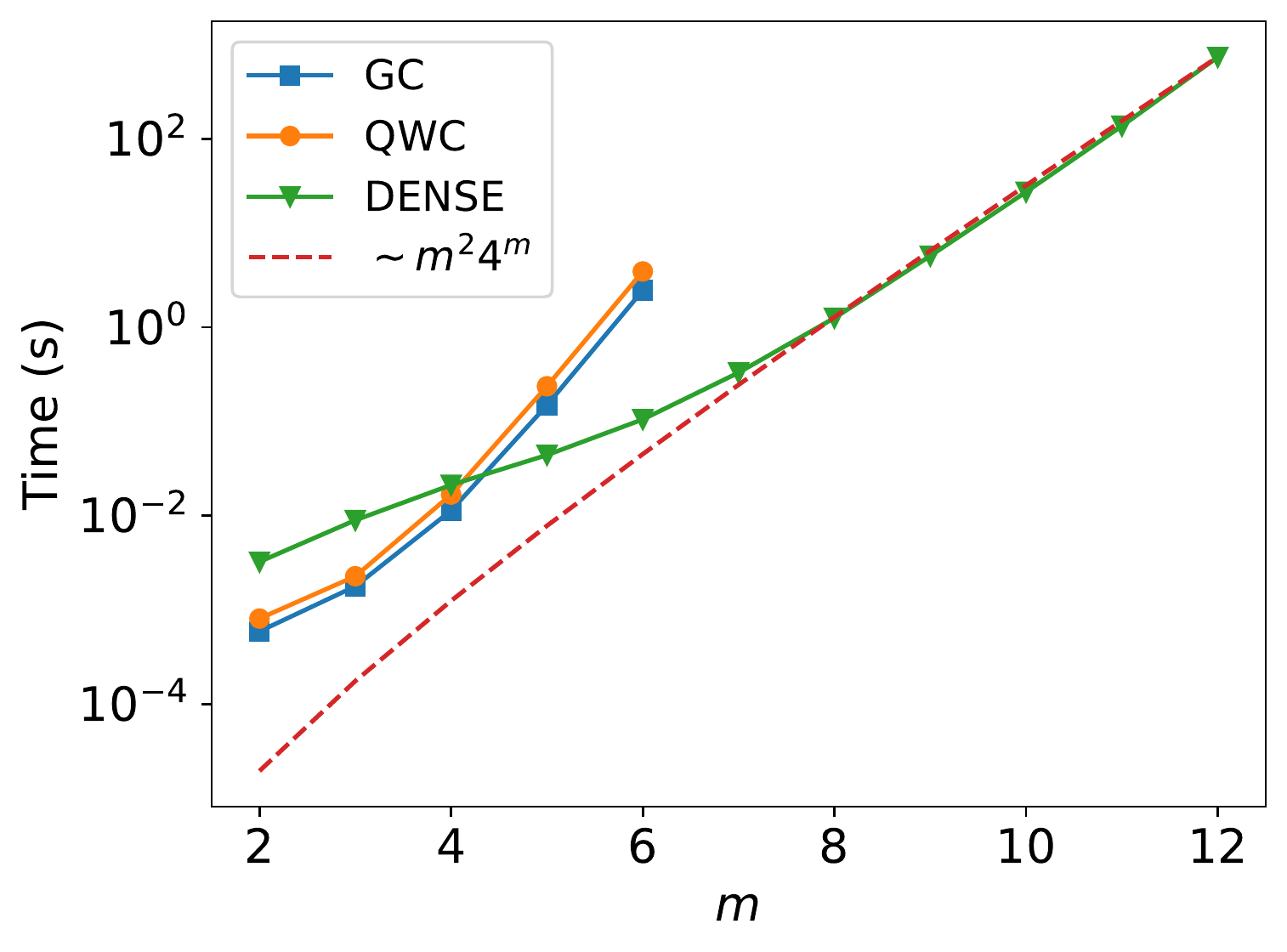}
\includegraphics[width=0.49\textwidth]{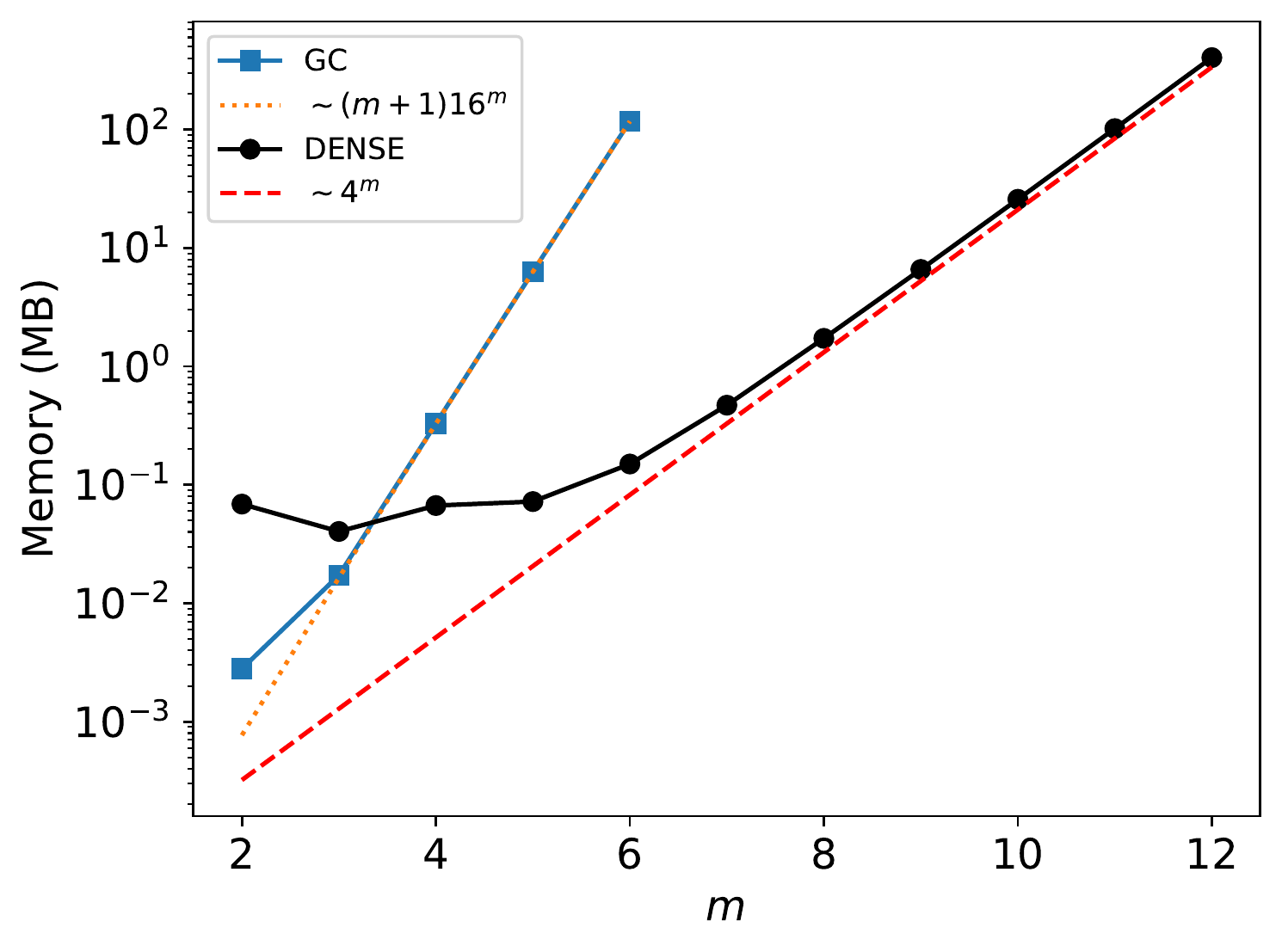}
\caption{(Left) Comparison of walltimes needed to group a random Hermitian operator using general (GC) and qubit-wise commuting (QWC) graph algorithms, compared to the DENSE method. An empirical curve
representing the large $m$ scaling of the DENSE method is shown as a red dashed line. (Right) Comparison of memory requirements for the GC and DENSE algorithms. For GC, we monitored only the memory needed to build the adjacency matrix. Empirical curves matching the large $m$ results are given as dotted/dashed lines.
}
\label{fig:timing2}
\end{figure*}

For fully dense matrices, graph algorithms generate a $4^m \times 4^m$ adjacency matrix representing the connectivity of the graph, and so we expect the memory use to scale at least as $16^m$. For DENSE, the solutions are generated by repeated application of an $m \times m$ $A$ matrix, and the memory required to store the solution increases as $4^m$. We empirically tested these expectations, with results shown in Fig.~\ref{fig:timing2} (right).
For the GC/QWC routines, we measured only the memory required for building the adjacency matrix in the {\tt \_noncommutation\_graph()} subroutine, while for DENSE the values represent total peak memory usage. Peak memory usage for GC/QWC was such that we could only access up to $m=6$ on our laptops, whereas for DENSE we could easily generate solutions to $m=12$ and beyond.

We also measured the walltimes required to generate solutions in the fully dense case.  Here we simply timed calls to either the {\tt group\_commuting()} or {\tt Pauli\_Organizer()} routines in {\tt Qiskit} and {\tt psfam}, respectively. For DENSE, solutions are generated by computing the orbit of generators produced by matrix powers of $A$, or alternatively by using the lookup procedure described in~\ref{sec:lookup}. In both cases we expect the timing to scale with the number of strings, up to sub-exponential corrections. This is shown for fully dense operators in Fig.~\ref{fig:timing2} (left). We find that for small $m$ the solution times for GC and DENSE are comparable, and that DENSE is faster for $m > 4$. Note that the DENSE method can also easily be made to run in parallel, which could further reduce walltimes.

\section{Conclusions} \label{sec: conclusion}
The Pauli grouping algorithm discussed in this paper can substantially reduce the time required to measure operators on quantum simulators and real devices. The algorithm completely sorts all Pauli strings on any fixed number of qubits $m$ into a minimal set of maximally-sized commuting families without any repeated strings. It is optimal for dense operators, with support over an order one fraction of the strings, and in this case it can reduce runtimes by a factor of $(2/3)^m$ relative to qubit-wise commuting groupings. The public string partitioning package {\tt psfam} and the Qiskit package {\tt dense\_ev} described in Sec~\ref{sec:code} have been tested on random Hamiltonians and a matrix quantum mechanics model relevant in high energy physics. We find timing improvements relative to Abelian grouping close to the theoretical limit and small impacts on precision, indicating that larger post-state rotation depths and correlated uncertainties are not significant effects at least for some problems of interest.

In its current form, this approach is mainly useful for problems involving dense operators and relatively small numbers of qubits. For sparse operators, the grouping makes no attempt to minimize the number of families that appear, and graph theoretic approaches are typically better. Combining the two methods could present interesting opportunities for future development. We emphasize, however, that in some cases there may not be a one-size-fits-all best method. Similar to integration, there may be a wealth of efficient techniques, analytic and numerical, which are effective against different classes of problems.  Also, the computational cost of measuring dense operators is still exponential in $m$, and therefore infeasible asymptotically. However, there are many interesting problems in physics and quantum information that involve dense operators, or operators with dense sub-blocks (as can arise in lattice gauge theories), that we would like to study with NISQ era devices. For such problems the reduction in measurement cost from, say, a thousand circuits to a hundred can be a significant consideration. 

A sample of interesting directions for further development include combining dense Pauli grouping with approximation methods to optimize precision for fixed resources; extending the method to optimize the grouping solution for operators of intermediate density; and studying applications to simulations of HEP models,  state tomography, and others. 
We hope to return to these problems in future work.

\section{Acknowledgements}
This work was supported in part by the U.S.
Department of Energy, Office of Science, Office of High Energy Physics under award number DE-SC0015655
and by its QuantISED program under an award for the Fermilab Theory Consortium ``Intersections of QIS
and Theoretical Particle Physics.” 
The authors thank Frank Harkins for Qiskit Ecosystem support, and IBM Quantum for assistance obtaining timestamped device calibration data.
We acknowledge the use of IBM Quantum~\cite{IBMQuantum} services for this work. The views expressed are those of the authors, and do not reflect the official policy or position of IBM or the IBM Quantum team.

\bibliographystyle{utphys}
\bibliography{citation}
\appendix

\section{Matrix algorithms}
\label{sec:matrix_algorithms}
In this appendix we describe the algorithmic construction of the various matrices used in Section \ref{sec: Algorithmic construction} to construct a suitable set of generating matrices $\{ A_1 ... A_{N-1} \}$. We will focus on exclusively $\mathbb{Z}_2$ valued matrices in this section. We start by constructing a companion matrix \cite{finitefields}. A companion matrix is constructed by using an irreducible polynomial $f(x) = \sum_{i=0}^{n} a_i x^i$ (which, for example, the galois.py package can produce \cite{Hostetter_Galois_2020}.) An irreducible polynomial is a polynomial over a finite field which can't be written as a product of polynomials in that field. The companion matrix is constructed as
\begin{equation}
    C = \begin{pmatrix}
        0 & 1 & 0 & ... & 0 \\
        ... & & 1 & &  \\
        0 & & 0 & & 1 \\
        -a_0 & -a_1 & ... & & -a_{m-1}
    \end{pmatrix}
\end{equation}
The matrix $D$ such that $CD = DC^T$ is:
\begin{equation}
    D = \begin{pmatrix}
        0 & 0 & 0 & ... & 0 & b_0 \\
        0 & 0 & 0 & ... & b_0 & b_1  \\
        0 & 0 & 0 & ... & b_1 & b_2 \\
        ... &  ... & ... & & ... & ... \\
        0 & b_0 & b_1 & ... & b_{m-3} & b_{m-2} \\
        b_0 & b_1 & b_2 & ... & b_{m-2} & b_{m-1}
    \end{pmatrix}
\end{equation}
 where the $b_i$ are defined as
\begin{equation}
    b_i = \sum_{k=0}^{i-1} a_{m-i+k} b_k
\end{equation}
with $b_0\equiv 1$. Multiplying the matrices, one finds:
\begin{equation}
        CD = \begin{pmatrix}
        0 & 0 & ... & b_0 & b_1 \\
        0 & 0 & ... & b_1 & b_2  \\
        0 & 0 & ... & b_2 & b_3 \\
        ... & ... & & ... & ... \\
        b_0 & b_1 & ... & b_{m-2} & b_{m-1}\\
        -a_{m-1}b_0 & -a_{m-1} b_1 + ... & ... & ... & ...
    \end{pmatrix}
\end{equation}
where the elements in the bottom row are
\begin{equation}
    (CD)_{m-1,i} = \sum_{k=0}^{(i+1)-1} a_{m-(i+1)+k} b_{k} = b_{i+1},
\end{equation}
so that $CD$ can be simplified to
\begin{equation}
    CD = \begin{pmatrix}
        0 & 0 & 0 & ... & b_0 & b_1 \\
        0 & 0 & 0 & ... & b_1 & b_2  \\
        0 & 0 & 0 & ... & b_2 & b_3 \\
        ... &  ... & ... & & ... & ... \\
        b_0 & b_1 & b_2 & ... & b_{m-2} & b_{m-1}\\
        b_1 & b_2 & b_3 & ... & b_{m-1} & b_{m}\\
    \end{pmatrix}.
\end{equation}
This can be verified by explicit matrix multiplication. The symmetry of $D$ and $CD$ implies $CD = DC^T$:
\begin{equation}
    C D = (CD)^T = D C^T.
\end{equation}

The next step of this process assumes that there is a 1 somewhere on the diagonal of $D$. If $m$ is odd, then there must be a 1 on the diagonal, because $b_0$ is on the diagonal. For $m$ even, the diagonals are populated by odd $b_i$. We will show by induction and contradiction that for even $m$, it is impossible that $b_i = 0$ for all odd $i$. Suppose that $b_1=0$. Then
\begin{align}
    b_1 = a_{m-1} b_0 &= 0 \nonumber\\
    \Rightarrow a_{m - 1} &= 0.
\end{align}
The inductive step is as follows. Take some odd $j>1$. Suppose $a_{m - i} = 0$ for all odd $i < j$. For these values, the indices of $a$ are odd.  Then
\begin{equation}
    b_j = 0 = \sum_{k = 0}^{j-1} a_{m - j + k} b_k 
\end{equation}
When $k$ is odd, $b_k = 0$. When $k$ is even and nonzero, $m-j-k$ is odd, so $a_{m-j-k} = 0$ by the assumption above. So the entire sum reduces to the $k = 0$ term, and we find that $a_{m-j}$ must vanish:
\begin{equation}
    b_j = a_{m - j} b_0 = a_{m-j} = 0.
\end{equation}
Thus, if all $b_i=0$ for odd $i$, we may iterate this process for increasing values of $j$ to find:
\begin{equation}
    a_i = \left\{ \begin{array}{cc}
        0  & i \text{ odd} \\
        c_{i/2}  & i \text{ even} \\
    \end{array} \right\} \\
\end{equation}
for some $c$. This corresponds to a perfect square of the polynomial $\sum_{i=0}^{m/2}c_ix^i$. But  $a$ describes an irreducible polynomial, which is a contradiction.  We conclude that it is impossible for $b_i=0$ for all odd $i$, so there must be some odd value of $i$ such that $b_i$ = 1, and this must appear on the diagonal of $D$, so there is a 1 on the diagonal of $D$. 

Next, we construct a matrix $\Lambda$ so that every diagonal entry of $M=\Lambda^T D \Lambda$ is unity. Choose an $a$  such that $D_{a,a} = 1$ and define a vector $v_{i} = 1 + D_{i,i}$.   Then construct
\begin{align}
    \Lambda_{i,j} = \delta_{i,j} + \delta_{i,a} v_j . 
\end{align}
Then 
\begin{align}
    (\Lambda^T D \Lambda)_{i,i} &=\sum_{j,k} (\delta_{j,i} + \delta_{j,a} v_i) D_{j,k} (\delta_{k,i} + \delta_{k,a} v_i) \nonumber\\
    &= D_{i,i} + D_{a,a} v_i^2 + D_{i,a} v_i + D_{a,i}v_i\nonumber\\
    &=  D_{i,i} + v_i \nonumber\\
    &= 1+  2D_{i,i}\nonumber\\
    & = 1
\end{align}
 gives an $M$ with 1 on every diagonal. (Note that $2D_{i,i} = 0$ because we are working in $\mathbb{Z}_2$.) 
 
 $M$ is invertible because it is a product of invertible matrices. It must also have a principal minor which is invertible, which we will prove next.
 
 Consider a symmetric principal minor $i$ of $M$, which is called $M[i]$. If $M[i]$ is not invertible then it has a eigenvector with zero eigenvalue:
\begin{align}
    \exists c \in \mathbb{Z}_2^{m}: \sum_{k \neq i} M[i]_{jk} c_k = 0 .
\end{align}

There is no eigenvector of zero eigenvalue for $M$, since $M$ is invertible. Let $c_i=0$. Then
\begin{align}
    (Mc)_{j\neq i} &= \sum_{k \neq i} M[i]_{jk} c_k + M_{ji}c_i = 0
\end{align}
and so it must be that $(Mc)_{j=i}=1$. Thus $Mc = u$, with $u$ defined by $u_j = \delta_{i,j}$. Therefore $c^T M c = c_i = 0$. 
We also have  $M^{-1} u = c$ and $u^T M^{-1} u = M^{-1}_{ii} = c_i =0$.

Now suppose that every principal minor is non-invertible. Then, for every $i$, we may construct a $\mathbb{Z}_2^{m}$-vector $c^i$, with $c^i_i=0$, which restricts to a $\mathbb{Z}_2^{m-1}$-eigenvector of $M[i]$ with zero eigenvalue. Therefore all diagonal elements of $M^{-1}$ vanish. Now sandwich $M^{-1}$ between any  vector $v$ in $\mathbb{Z}_2^{m}$:
\begin{equation}
    v^T M^{-1} v = \sum_{i,j} v_i  M^{-1}_{ij} v_j
\end{equation}
In the sum, every off-diagonal term of $M^{-1}$ appears with its transpose. 
Since $M$ is symmetric and we are working in $\mathbb{Z}_2$ these terms cancel. So the only terms remaining are:
\begin{equation}
    v^T M^{-1} v = \sum v_i^2 M^{-1}_{ii} = 0.
\end{equation}
Thus the equation $v^T M^{-1} v = 0$ holds for every $v$. In particular, for $v_j\equiv M_{ji} $ for any $i$, we find

\begin{align}
   0 = v^T M^{-1} v = (M^T)_{ij} M^{-1}_{jk} M_{ki} = M_{i,i}^T = 1 .
\end{align}

\noindent This is a contradiction, so $M$ has at least one principal minor $i$ which is invertible and symmetric. 

Assume $M=LL^T$ and $M[i] = L[i] L[i]^T$ (we will use induction to prove these decompositions momentarily). Then:
\begin{align}
    M &= 
    \begin{pmatrix}
        L[i] L[i]^T & \eta \\
        \eta^T & 1
    \end{pmatrix} \nonumber\\ 
    &= \begin{pmatrix}
        L[i] & 0 \\
        (L[i]^{-1} \eta)^T & d_0
    \end{pmatrix} \begin{pmatrix}
        L[i]^T & L[i]^{-1} \eta \\
        0 & d_0
    \end{pmatrix}
\end{align}
The multiplication of the last row and column gives $1 = d_0^2 + \sum_j (L_i^{-1} \eta)_j^2$. If $d_0 = 0$, then $L$ is not invertible. This cannot be the case. If $L$ is not invertible, then there is a vector $v$ such that $L^T v = 0$. Then $L L^T v = 0$, which means that $M$ is not invertible. But we have already seen that $M$ is invertible, so $d_0 = 1$.

The induction process works as follows. $M$ has an invertible principal minor, which also has an invertible principal minor, until the principal minor is $[1] = [1] [1]^T$. Each principal minor can be written as $L L^T$, so the matrix which has $L L^T$ as its principal minor can also be written as $L L^T$, until $M = L L^T$.

The algorithm which completes this process must take the inverse of $L$, which has $O(m^3)$ complexity. The entire algorithm should then have complexity $O(\sum_{i=0}^m i^3) = O(m^4)$.

\section{Circuit representation}
\label{sec:circuit_rep}

\begin{figure*}[h!]
    \centering
    \includegraphics[width = 0.45\linewidth]{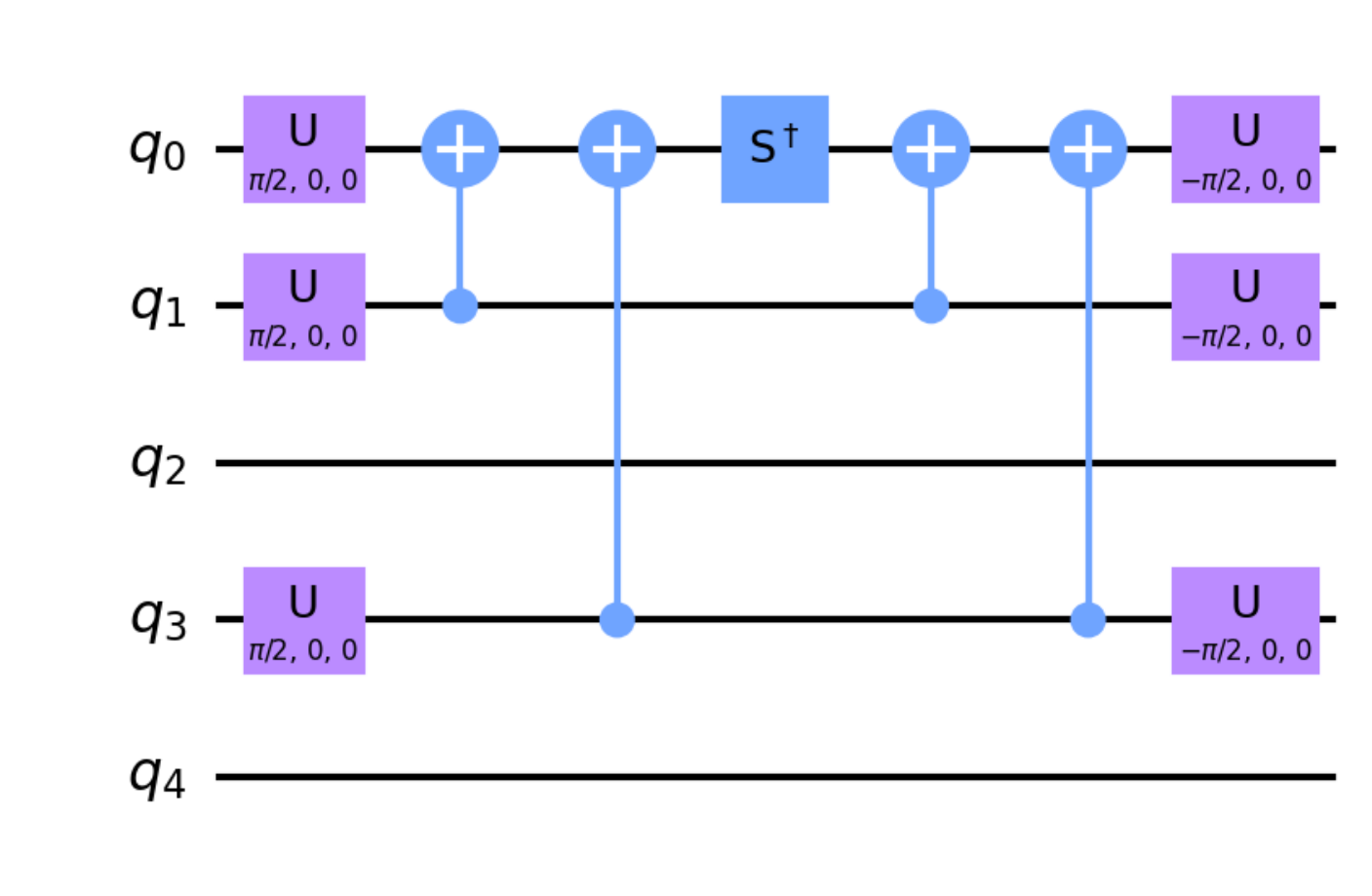}~~
    \includegraphics[width = 0.45\linewidth]{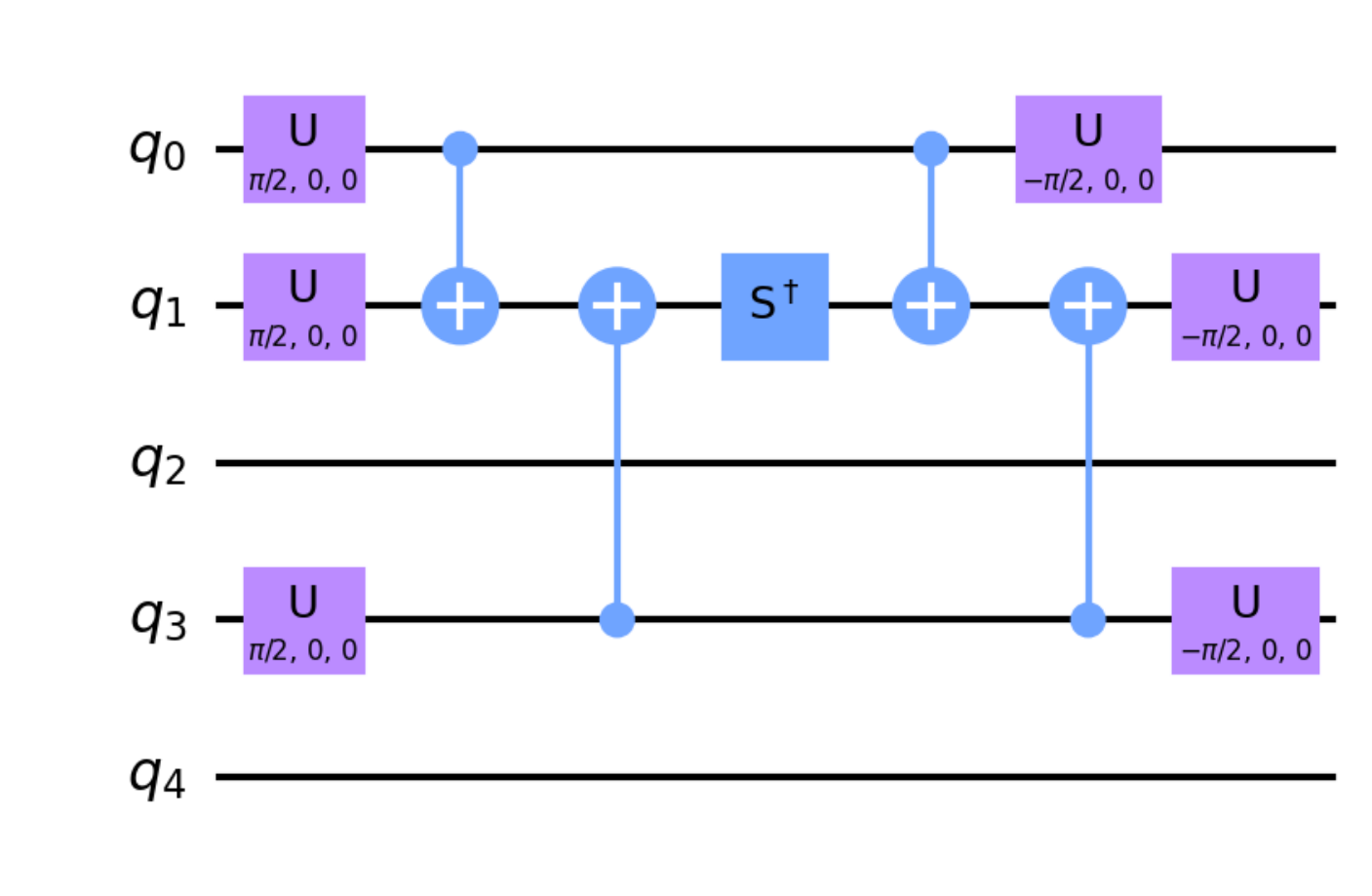}
    \caption{2 possible representations of the quantum circuit corresponding to $\exp(\frac{i \pi}{4} IXIXX)$.}
    \label{fig:example_circuit}
\end{figure*}

This appendix details the circuit representation of $U = \exp(i \frac{\pi}{4} \sum_k x_k)$. First, a circuit representation of $e^{\frac{i \pi}{4} x_k}$ is needed. We decompose the factors in $U$ as
\begin{align}
    e^{\frac{i \pi}{4} x_k} = \frac{1}{\sqrt{2}} (1+i x_k) = \frac{1}{\sqrt{2}}U_Y^\dagger (1+i z_k) U_Y, \\
    U_Y = \prod_i^m e^{ \frac{i \pi}{4} y_{2^i}} 
\end{align}
    so that we find
\begin{align}
    e^{\frac{i \pi}{4} \sum x_k} = U_Y^\dagger e^{ \frac{i \pi}{4} \sum z_k} U_Y.
\end{align}
In the definition of $U_Y$,  $y_{2^i}$ represents a string with a single $Y$ at the $i$th position, and $I$ everywhere else; as a tensor product, it is
\begin{equation}
    e^{\frac{i \pi}{4} y_{2^i}} = \mathds{1} \otimes ... \frac{1 + i \sigma_y}{\sqrt{2}} ... \otimes \mathds{1}.
\end{equation}
Since each $e^{\frac{i \pi}{4} y_{2^i}}$ is a tensor product, computing $U_Y$ is the same as computing a one qubit gate for $\frac{1 + i \sigma_y}{\sqrt{2}}$ and applying it to every qubit. This one qubit gate is a U gate with parameters $\theta = -\pi, \lambda = \phi = 0$, so $U_Y$ is a product of U$(-\pi,0,0)$ gates on each qubit.

The operator $e^{\frac{i \pi}{4} z_k}$ can be expressed as a quantum circuit as follows. First, find the set of qubits on which the string $z_k$ contains a $\sigma_z$ in its tensor product. Select one  qubit $Q$ from the set and apply $CX$ gates between $Q$ and all the rest in the set. This measures the parity of the set. Then apply an $S$ gate to $Q$, which is equal to $\sqrt{Z}$ up to a phase. Finally undo the $CX$ gates.  See Figure \ref{fig:example_circuit} for an example. 

We have so far proven that any of our unitary transformations can be expressed as a quantum circuit, except for the $z$ and $x$ families. The $x$ family is diagonalized by $U_Y$ and the $z$ family is already diagonal:
\begin{equation}
    U_Y Z U_Y^\dagger = \frac{1}{2}(1 + iY) Z(1 - iY) = i (-iZ) = +X.
\end{equation}
This can be repeated for each qubit to diagonalize any $x$ string.

To calculate the expectation value of a Hamiltonian, $\langle \psi | H | \psi \rangle$, follow the following procedure. First decompose the Hamiltonian into Pauli strings. As in the main text we label the strings based on  families:
\begin{equation}
    T_{i,j} = \left\{ \begin{array}{cc}
        S_{i,P^{(j)}(i)} & \text{if } 0 <  j < N \\
        z_i & \text{if } j = 0\\
        x_i & \text{if } j = N
    \end{array} \right\}.
\end{equation}
Here $P^{(j)}$ represents the permutation associated with the $j$th family, and $T_{i,j}$ is the $i$th \emph{string} of the $j$th \emph{family}. We may decompose $H$ as
\begin{equation}
    \langle H \rangle = \alpha_{0,0} + \sum_{i=1}^{N-1} \sum_{j = 0}^N \alpha_{i,j} \langle T_{i,j} \rangle
\end{equation}
There is a coefficient for each measurement including the identity, whose decomposition we label $\alpha_{0,0}$, placing it with the $z$ family. 

Let $|\chi_k \rangle$ represent the $k^{\rm th}$ \emph{state} in the computational basis. (Recall we are ordering this basis via binary encoding, where, e.g., state $6$ is $| 110 \rangle $ or $| \uparrow \uparrow \downarrow \rangle$.) Let $| \phi_k^{(j)} \rangle $ represent the $k^{\rm th}$ eigenvector of the family $j$. If $U_j$ diagonalizes family $j$, then $U_j | \phi_k^{(j)} \rangle  = | \chi_k \rangle $. Then:
\begin{align}
    \langle T_{i,j} \rangle = \sum_{k=0}^{N-1} \langle \psi | T_{i,j} | \phi_k^{(j)} \rangle \langle \phi_k^{(j)} | \psi \rangle \nonumber \\
    = \sum_{k=0}^{N-1} \lambda_{i,k} \langle \psi| \phi_k^{(j)} \rangle \langle \phi_k^{(j)} | \psi \rangle \nonumber \\
    = \sum_{k=0}^{N-1} \lambda_{i,k} \langle \psi|U_j^\dagger| \chi_k \rangle \langle \chi_k |U_j | \psi \rangle \nonumber \\
    = \sum_{k=0}^{N-1} \lambda_{i,k} | \langle \chi_k | U_j | \psi \rangle |^2
\end{align}

The value $ | \langle \chi_k | U_j | \psi \rangle |^2$ represents the fraction of measurements on $U_j$ which are measured in the $k$ state. Placing this into the equation for the Hamiltonian:
\begin{multline}
    \langle H \rangle = \sum_{i = 1}^{N-1} \sum_{j = 1}^N \sum_{k = 0}^{N-1} \alpha_{i, j} \lambda_{i,k} | \langle \chi_{k} | U_j | \psi \rangle |^2 \\
    + \sum_{i = 0}^{N-1} \sum_{k = 0}^{N-1} \alpha_{i,0} \lambda_{i,k} | \langle \chi_{k} | \psi \rangle |^2
\end{multline}
For clarity we can factor this expression into the measurements times a set of coefficients. Each family will have $N$ coefficients, representing the $N$ common eigenstates:
\begin{equation}
    \langle H \rangle = \sum_{j = 0, k = 0}^{N} c_{j,k} M_{j,k} 
\end{equation}
Here $M_{j,k}$ represents the fraction of measurements of family $j$ in state $k$. The coefficients are
\begin{equation}
    c_{j,k} = \left\{ \begin{array}{cc}
        \sum_{i=1}^{N-1} \alpha_{i,j}\lambda_{i,k} & \text{if } 0 <  j \leq N \\
        \sum_{i=0}^{N-1} \alpha_{i,0}\lambda_{i,k} & \text{if } j = 0\\
    \end{array} \right\}.
\end{equation}

\section{Device characteristics}
\label{sec:device}
\begin{figure}[htb]
\includegraphics[width=0.20\textwidth]{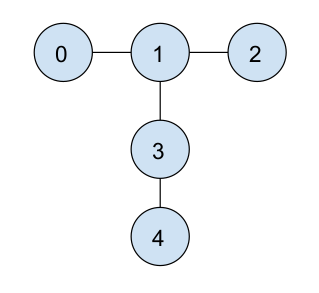}
\caption{Device layout for {\tt ibmq\_quito}. Circles represent qubits and solid lines represent connectivity between qubits.
\label{fig:layout}}
\end{figure}

The demonstrations in Sec.~\ref{sec:runtimes} were carried out on {\tt ibmq\_quito}. The device layout is shown in
Fig.~\ref{fig:layout}.
Device characteristics from the time of the demonstrations are tabulated below.

\begin{table}[h]
    \centering
    \begin{tabular}{ |c|c|c|c|c|c|}
     \hline
        Qubit & T1(us)  & T2(us) & Freq.(GHz) & Anharmonicity(GHz) & Readout error  \\
    \hline
        0 & 69.885 & 109.664& 5.3006 & -0.33148 & 0.0405 \\
    \hline
        1 & 73.686 & 109.025& 5.0805 &  -0.31924 & 0.0419 \\
    \hline
        2 & 91.183 & 94.412& 5.3221&-0.33231 &0.0647 \\
    \hline
        3 & 65.835 & 17.463 &  5.1635& -0.33508 & 0.0547  \\
    \hline
        4 & 20.609 & 39.990 &5.0522 & -0.31926 & 0.0436  \\
    \hline
         
    \end{tabular}

    \vspace{1cm}

      \begin{tabular}{ |c|c|c|c|c|c|}
      \hline
        Qubit & Readout length(ns) & Prob meas0 prep 1 & Prob meas1 prep 0 & ID error & sx error  \\
        \hline
        0 & 5351.11 &0.0596  & 0.0214 & 0.00026747 & 0.00026747  \\
        \hline
        1 & 5351.11 &0.0636 & 0.0202& 0.00027251& 0.00027251\\
        \hline
        2 & 5351.11 & 0.0778& 0.0516 & 0.00023647&0.00023647   \\
        \hline
        3 & 5351.11 & 0.0870&0.0224 &0.0013701& 0.0013701  \\
        \hline
        4 & 5351.11 &0.0626 &0.0246 & 0.00047614 & 0.00047614 \\
        \hline
        \end{tabular}

    \vspace{1cm}
CNOT gate errors
    \begin{tabular}{ |c|c|c|c|c|c|}
    \hline
        Qubit & 0 & 1 & 2  & 3 & 4  \\
        \hline
        0 & - & 0.0074208 & - & - &  -  \\
        \hline
        1 & 0.0074208& - & 0.0067878&0.0182670  &-    \\
        \hline
        2 & - & 0.0067878& - & -& -  \\
        \hline
        3 & - & 0.0182670& - & - & 0.0213555   \\
        \hline
        4 & - & - & - &  0.0213555& -  \\
        \hline
        \end{tabular}
    
    \caption{Device data for {\tt ibmq\_quito}, measured on April 03, 2023. The first two tables give the collected single-qubit data, while the third gives measured CNOT gate errors between connected device qubits.}
    \label{tab:device}
\end{table}

\end{document}